\documentclass{pas}

\usepackage{multirow}
\usepackage{newtxtext}
\usepackage{newtxmath}
\usepackage{tabularx}
\usepackage[T1]{fontenc}

\usepackage{graphicx}	% Including figure files
\usepackage{amsmath}	% Advanced maths commands
\usepackage{longtable}
\usepackage[table]{xcolor}
\RequirePackage{etex}

\begin{document}

\lefttitle{Fifty years of primordial helium abundances}
\righttitle{Nabil Husain, Richard de Grijs and Giuseppe Bono}

\jnlPage{1}{20}
\jnlDoiYr{2026}
\doival{10.1017/pasa.xxxx.xx}

\articletitt{Research Paper}

\title{Fifty years of primordial helium abundances: A statistical reanalysis}

\author{\sn{Nabil} \gn{Husain}$^{1}$, \sn{Richard} \gn{de Grijs}$^{2,3,4}$ and \sn{Giuseppe} \gn{Bono}$^{5,6}$}

\affil{$^{1}$Indian Institute of Science Education and Research (IISER) Mohali, Knowledge City, Sector 81, Manauli, PO, Sahibzada Ajit Singh Nagar, Punjab 140306, India\\
$^{2}$School of Mathematical and Physical Sciences, Macquarie University, Balaclava Road, Sydney, NSW 2109, Australia\\
$^{3}$Astrophysics and Space Technologies Research Centre, Macquarie University, Balaclava Road, Sydney, NSW 2109, Australia\\
$^{4}$International Space Science Institute--Beijing, 1 Nanertiao, Zhongguancun, Hai Dian District, Beijing 100190, China\\
$^{5}$Dipartimento di Fisica, Universit\`a di Roma Tor Vergata, via Della Ricerca Scientifica 1, I-00133, Rome, Italy\\
$^{6}$INAF, Rome Astronomical Observatory, via Frascati 33, I-00078 Monte Porzio Catone, Italy}

\corresp{Richard de Grijs, Email: richard.de-grijs@mq.edu.au}

\citeauth{Nabil Husain, Richard de Grijs and Giuseppe Bono, Fifty years of primordial helium abundances: A statistical reanalysis. {\it Publications of the Astronomical Society of Australia} {\bf 00}, 1--23 (2026). https://doi.org/10.1017/pasa.xxxx.xx}

\history{(Received xx xx xxxx; revised xx xx xxxx; accepted xx xx xxxx)}

\begin{abstract}
The primordial helium mass fraction, $Y_\mathrm{p}$, is a key observational pillar of Big Bang nucleosynthesis and a sensitive probe of early-Universe physics. Over the past several decades, numerous observational $Y_\mathrm{p}$ determinations have been published using a wide range of astrophysical tracers and cosmological techniques. Although recent measurements exhibit striking convergence and increasingly small uncertainties, the statistical and historical context of this consensus has not been examined systematically. Here, we compile and analyse a comprehensive dataset of observational $Y_\mathrm{p}$ determinations published between the late-1960s and 2022. The final sample comprises 143 reported values spanning multiple tracers. We find clear evidence for long-term convergence in published $Y_\mathrm{p}$ values, punctuated by statistically significant change points in the mid-2000s and early 2010s. Careful examination reveals that many extragalactic H{\sc ii}-region determinations are not fully independent, relying on re-analyses or partial reuse of a limited number of observational datasets. This reduces the effective number of independent constraints and provides important context for interpreting the precision of recent results. Our findings do not challenge the overall consistency of modern $Y_\mathrm{p}$ determinations with standard cosmology, but they underscore the importance of accounting for data dependence, methodological homogeneity and historical evolution when synthesising measurements.
\end{abstract}


\maketitle

\section{Introduction}

The primordial mass fraction of helium, $Y_\mathrm{p}$, occupies a central place in modern cosmology. As one of the key light-element abundances produced during Big Bang nucleosynthesis (BBN), it provides a sensitive and largely independent probe of the physical conditions in the early Universe, complementing constraints derived from deuterium, lithium and the cosmic microwave background \citep[CMB;][]{1966ApJ...146..542P, 1967ApJ...148....3W, 2016RvMP...88a5004C}. Early estimates of the primordial helium abundance, beginning with the first extragalactic emission-line analyses in the late 1960s \citep{1969BOTT....5....3P}, already recognised both the promise and the systematic challenges of the method. In the context of the standard cosmological model, the value of $Y_\mathrm{p}$ depends primarily on the baryon-to-photon ratio, the neutron lifetime and the effective number of relativistic species \citep{2007ARNPS..57..463S, 2018PhR...754....1P}. Consequently, increasingly precise determinations of $Y_\mathrm{p}$ have long been viewed as a powerful means of testing the internal consistency of the prevailing $\Lambda$CDM cosmology and of searching for physics beyond the Standard Model \citep{2020JCAP...03..010F}.

A similar set of considerations applies to stellar diagnostics used to estimate the primordial helium abundance. Since the seminal work of \citet{1968Natur.220..143I}, who introduced the so-called $R$ parameter---defined as the ratio of the number of horizontal-branch stars to the number of red-giant-branch stars brighter than the horizontal branch at the level of the RR~Lyrae instability strip---a variety of stellar indicators have been proposed to constrain helium abundances in Galactic globular clusters. Many of these studies rely on broadly similar photometric datasets, often derived from the same ground-based imaging campaigns \citep[e.g.,][]{1986A&AS...66...79B, 2004AandA...420..911S} or from common {\sl Hubble Space Telescope} observations \citep[e.g.,][]{2000ApJ...538..289Z}. As a result, nominally independent determinations may in practice share significant underlying data and modelling assumptions, paralleling the situation encountered for extragalactic H{\sc ii}-region studies. For a more detailed discussion of helium abundance diagnostics based on stellar evolutionary indicators, we refer the reader to \citet{2000MNRAS.313..571S} and \citet{2011PASP..123..879T}.

Note that throughout this paper we adopt the traditional notation of the primordial helium mass fraction, $Y_\mathrm{p}$, since this quantity has historically dominated the observational and cosmological literature analysed here. We note, however, that modern studies increasingly report the primordial helium \textit{number} abundance, $y_\mathrm{P} \equiv n(\mathrm{^4He})/n(\mathrm{H})$, particularly in the context of H{\sc ii}-region spectroscopy, CMB analyses and BBN calculations, where the underlying observables are naturally expressed in terms of number densities rather than mass fractions. In practice, many modern determinations infer helium number abundances first and subsequently convert these to $Y_\mathrm{p}$ for comparison with the historical literature.

Unlike deuterium, whose primordial abundance is inferred from a relatively small number of high-redshift absorption systems \citep[e.g.,][]{2018ApJ...855..102C}, estimates of $Y_\mathrm{p}$ are typically obtained by extrapolating observations of helium emission lines in low-metallicity extragalactic H{\sc ii} regions to zero metallicity \citep[e.g.,][]{1969BOTT....5....3P, 1992MNRAS.255..325P}. This approach has benefited from major advances in observational capabilities, atomic data and photo-ionisation modelling \citep[e.g.,][]{1999ApJ...514..307B, 2012MNRAS.425L..28P}. At the same time, it is also subject to a complex web of systematic uncertainties, including temperature and density structure, underlying stellar absorption, ionisation corrections, collisional excitation, reddening and the treatment of chemical evolution \citep[e.g.,][]{2004ApJ...617...29O, 2007ApJ...666..636P}. As a result, published determinations of $Y_\mathrm{p}$ have exhibited a non-negligible dispersion, even among studies using ostensibly similar datasets and methodologies.

Although all published determinations aim to constrain the same physical quantity, the routes by which different observations lead to an inferred value of $Y_\mathrm{p}$ vary substantially. In astrophysical measurements, $Y_\mathrm{p}$ is typically not observed directly but inferred from present-day helium abundances in specific environments, combined with assumptions about chemical evolution and stellar processing. By contrast, cosmological determinations extract $Y_\mathrm{p}$ indirectly from its imprint on early-Universe physics, most notably through its effect on the recombination history encoded in the CMB anisotropies. As a result, different tracers probe different epochs, physical processes and sources of uncertainty even though they ultimately target the same primordial parameter. Understanding these inference pathways is therefore essential when interpreting patterns in the literature and assessing the degree to which different measurements can be considered independent or comparable.

Over the past three decades, a substantial literature has emerged reporting progressively more precise values of $Y_\mathrm{p}$. A useful early overview of the historical evolution of published $Y_\mathrm{p}$ determinations was presented by \citet[][his Figure~10]{2007ARNPS..57..463S}, although without the statistical and bibliographic focus adopted here. Reported values are often accompanied by claims of concordance with CMB-derived baryon densities \citep[e.g.,][]{2010ApJ...710L..67I, 2015JCAP...07..011A} or, conversely, mild tensions that have been interpreted as hints of new physics \citep[e.g.,][]{Steigman2012}. This progressive tightening of error bars and apparent convergence towards a canonical value raises an important and rarely examined question: to what extent does the published record of $Y_\mathrm{p}$ measurements faithfully reflect the underlying observational uncertainties, and to what extent might it be shaped by sociological or methodological effects that influence which results are published, emphasised or cited? Similar concerns have been raised in other areas of observational astrophysics, where cumulative measurements of a single parameter exhibit progressive tightening and apparent convergence over time, even when derived from heterogeneous datasets and methods.

Concerns of this type are commonly discussed under the umbrella of `publication bias', a term that encompasses a range of effects whereby results that are perceived as more interesting, more precise or more consistent with prevailing expectations are preferentially disseminated \citep[e.g.,][]{Ioannidis2005, Fanelli2012}. Throughout this paper, we use the term `publication bias' in a broad and descriptive sense, referring to systematic patterns in the published literature that may arise from shared assumptions, community expectations or methodological conventions rather than from any deliberate selection or suppression of results. In the astronomy and astrophysics literature, explicit quantitative studies of publication bias remain comparatively rare, despite the fact that many key parameters---distances, cosmological constants and fundamental scaling relations among them---are derived from heterogeneous datasets and complex analyses that are vulnerable to both conscious and unconscious selection effects \citep[e.g.,][]{Henrion2013}.

In a recent series of papers, we explored the possible presence and consequences of publication bias in published distance estimates to galaxies in the Local Group and beyond \citep{2014AJ....147..122D, 2014AJ....148...17D, 2015AJ....149..179D, 2016ApJS..227....5D, 2017ApJS..232...22D, 2020ApJS..246....3D, 2020ApJS..248....6D}. By compiling comprehensive databases of distance measurements and examining their temporal evolution, dispersion and reported uncertainties, we demonstrated that the literature record exhibits statistically significant signatures of convergence that are difficult to reconcile with purely random measurement errors, but which can more likely be attributed to the expected evolution in precision as better data become available and better methods are developed. In particular, we found evidence that quoted uncertainties often underestimate the true scatter among independent determinations, and that later measurements tend to cluster more tightly around community-accepted benchmark values than would be expected from methodological improvements alone. Importantly, these effects emerged despite the absence of any centralised experimental programme or formal combination procedure, highlighting how convergence can arise organically within a research community.

The present paper extends this line of inquiry to the primordial helium abundance. There are several compelling reasons to consider $Y_\mathrm{p}$ as a test case for potential publication bias. First, $Y_\mathrm{p}$ is a single scalar parameter of high theoretical importance, making it an obvious focal point for comparisons with external constraints from BBN theory and the CMB \citep[e.g.,][]{2020AandA...641A...6P}. Second, its determination relies on a chain of assumptions and corrections that are difficult to validate independently, increasing the scope for correlated systematic effects across different studies. Third, the literature spans several decades, during which both observational techniques and the cosmological context in which results are interpreted have evolved substantially.

Unlike previous reviews, which have focused primarily on refining individual determinations or reconciling subsets of measurements, the present paper treats the published literature itself as an object of study. By assembling a comprehensive historical database of $Y_\mathrm{p}$ determinations and analysing their temporal, methodological and statistical structure, we aim to characterise how consensus has emerged in the field and to assess whether patterns in the published values are consistent with purely methodological progress or may also reflect broader contextual influences. In other words, our goal in this paper is not to reassess the astrophysical methods used to derive $Y_\mathrm{p}$, nor to advocate a new `best' value of the primordial helium abundance. Instead, we ask an equally important question: does the ensemble of published $Y_\mathrm{p}$ values exhibit statistical patterns that are indicative of publication bias or other non-random influences on the reporting of results? By analysing a comprehensive database of $Y_\mathrm{p}$ determinations drawn from the peer-reviewed literature, we seek to characterise the evolution of reported central values and uncertainties, to compare the quoted error bars with the observed dispersion among independent measurements and to assess whether the degree of apparent convergence is consistent with expectations based on methodological progress alone.

This paper is organised as follows. In Section~\ref{data.sec} we describe the construction of our database of published $Y_\mathrm{p}$ determinations and the criteria used for inclusion. Section~\ref{methods.sec} presents a statistical analysis of the temporal evolution of reported values and uncertainties, including tests for excess clustering and variance suppression. In Section~\ref{discussion.sec} we discuss the implications of our findings for the interpretation of $Y_\mathrm{p}$ measurements in cosmology and place them in the broader context of publication bias in astrophysical parameter estimation. Our main conclusions are summarised in Section~\ref{concl.sec}.

\section{Data}
\label{data.sec}

This study is based on a dataset of primordial helium mass-fraction ($Y_\mathrm{p}$) determinations published between the late-1960s and 2022 \citep{2022ApJ...941..167M}. The guiding principle in assembling the dataset was completeness: the goal was to capture essentially all published observational estimates of $Y_\mathrm{p}$ across different tracers and methods. The concept of a primordial helium mass fraction emerged gradually in the literature, and early observational estimates were often framed in terms of helium abundances in specific astrophysical environments rather than as direct measurements of a cosmological parameter. At the same time, theoretical predictions based on BBN were beginning to establish an expected range for the primordial helium abundance \citep[e.g.,][]{1966ApJ...146..542P, 1967ApJ...148....3W}. The first explicit determinations of a primordial helium abundance appeared in the late 1960s and early 1970s in the context of these emerging BBN models and increasingly detailed spectroscopic studies of H{\sc ii} regions. These early values were necessarily exploratory, relying on limited data and simplified treatments of systematic effects, and they should be understood as establishing the feasibility of the approach rather than as precise measurements in the modern sense. Whereas these early studies differ substantially in methodology from later work, they define the starting point of the time series analysed here and illustrate the extent to which both observational capabilities and theoretical expectations have evolved over the subsequent decades. Although early determinations were subject to large uncertainties and simplified modelling assumptions, they already established the basic expectation that $Y_\mathrm{p}$ lies near $\sim 0.23$--0.25, a range that continues to frame modern discussions.

\subsection{Literature Compilation}

\begin{figure*}
 \includegraphics[width=17cm]{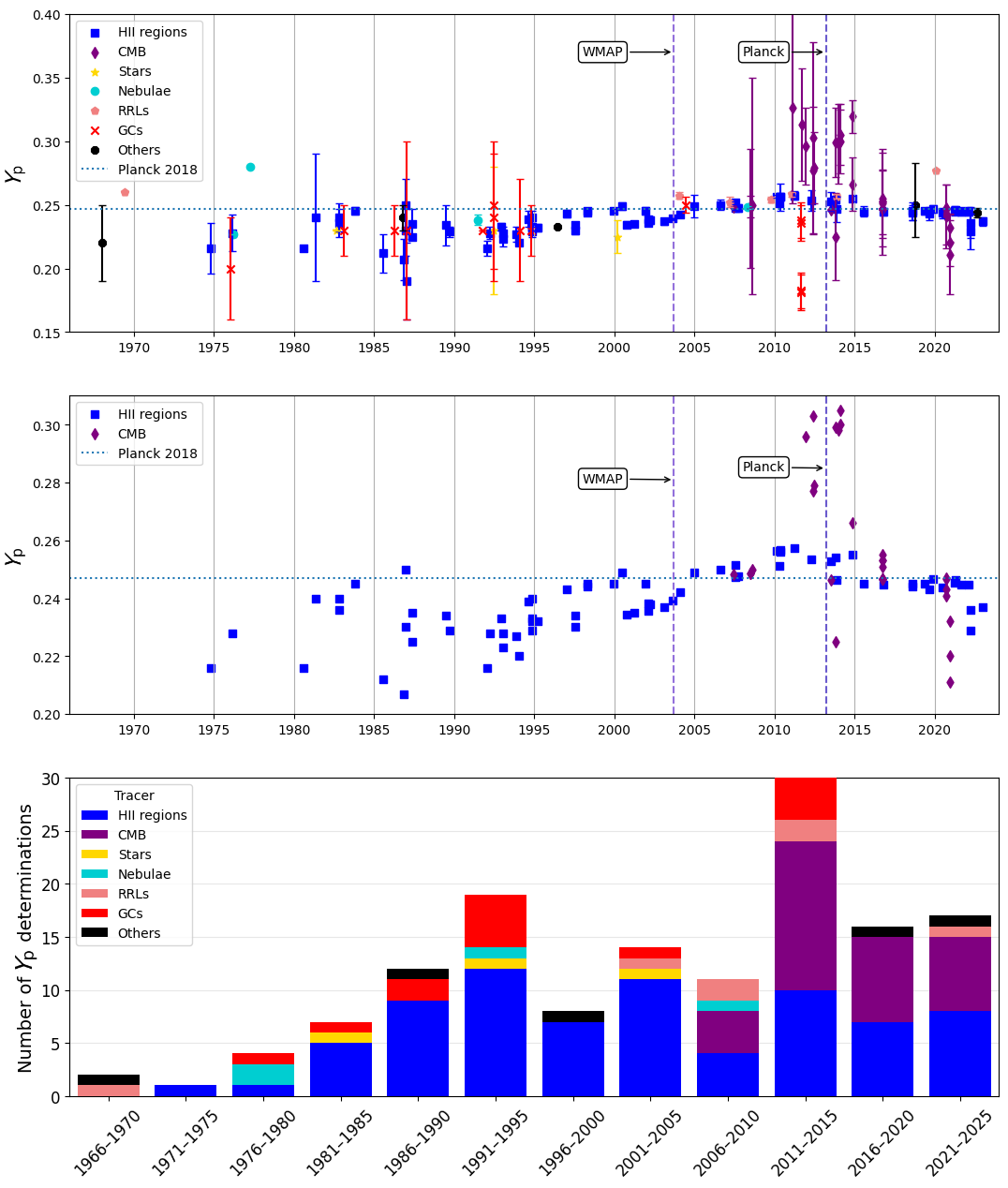}
 \caption{(Top) Published $Y_\mathrm{p}$ values with their uncertainties (where available) as a function of publication date, colour-coded by tracer. The {\sl Planck} 2018 value, $Y_\mathrm{p} = 0.2463$ (horizontal dotted line) is included for visual benchmarking. CMB: Cosmic Microwave Background; RRLs: Radio recombination lines; GCs: Globular clusters. (Middle) As the top panel but for the H{\sc ii}-region- and CMB-based analyses only. The publication dates of the {\sl WMAP3} (2004) and {\sl Planck} (2013) $Y_\mathrm{p}$ values are highlighted by vertical dashed lines. (Bottom) Numbers of $Y_\mathrm{p}$ determinations published within successive five-year intervals, grouped by observational tracer.}
 \label{fulldata.fig}
\end{figure*}

Candidate papers were identified through extensive searches of the NASA Astrophysics Data System (ADS). An initial set of queries combined the terms `primordial helium', `$Y_\mathrm{p}$' and `observational'. However, it soon became clear that some studies, especially from the nascent years of the field, did not explicitly use the standard notation $Y_\mathrm{p}$ or described their results as `primordial' helium abundance in titles or abstracts, hence we broadened the scope by omitting both `$Y_\mathrm{p}$' and `primordial' in subsequent searches. Analyses based on specific observational tracers also often adopted terms characteristic of their respective subfields like the $R$ parameter in the analyses of stellar populations in globular clusters or the use of data from {\sl Planck} in CMB-based work. Hence, tracer-specific queries (`r parameter’, globular’, `Planck’) were employed to include those. The resulting list was cross-checked against major review articles \citep[e.g.,][]{2012MSAIS..22..164S, 2021MNRAS.502.3045K}. The dataset was expanded further through citation chaining, i.e., by examining both references cited by relevant papers and later papers that cited them. In total, close to 2000 articles were vetted to arrive at the final list of published $Y_\mathrm{p}$ values. Some studies, particularly those based on metal-poor stars or globular clusters, found primordial helium fractions which were said to be `generally consistent' with $Y_\mathrm{p}$; however, only papers explicitly reporting a value of $Y_\mathrm{p}$ were included in the final dataset. Figure \ref{fulldata.fig} (top) provides an overview of the full dataset, colour-coded by tracer, as a function of time. To illustrate the growth and changing composition of the field over time, we also constructed binned histograms showing the number of reported $Y_\mathrm{p}$ determinations in successive five-year intervals, colour-coded by tracer class: see Figure \ref{fulldata.fig} (bottom). These distributions provide an overview of the publication rate and the emergence of new observational techniques used to determine $Y_\mathrm{p}$.

Despite extensive efforts to ensure completeness, we cannot exclude the possibility that some relevant determinations have been missed, particularly in cases where $Y_\mathrm{p}$ is reported indirectly or not explicitly labelled as such. However, given the breadth of the search strategy and our cross-checking against major review articles, any such omissions are unlikely to introduce systematic bias.

\subsection{Extraction}
From each paper, the following information was recorded:  
\begin{itemize}
    \item publication date (month and year);
    \item the quoted $Y_\mathrm{p}$ value(s);
    \item the associated statistical uncertainties, retaining asymmetry where reported;
    \item the tracer or method employed;
    \item the dataset (or spectra) used; and
    \item the full bibliographic reference.
\end{itemize}

Where multiple values were reported within a single study, these typically reflect alternative modelling assumptions or subsamples; such values are treated as distinct determinations but are tracked at the paper level to allow for later tests of non-independence. Of the 143 distinct $Y_\mathrm{p}$ values identified, two were excluded from further analysis: one determination with uncertainties comparable to the full physically plausible range \citep{1984AJ.....89..433M} and one result reported as an upper limit rather than a measurement \citep{2009ApJS..180..306D}. No additional trimming or selection was applied. The final dataset comprises 141 observationally inferred determinations of $Y_\mathrm{p}$, excluding purely theoretical or BBN-predicted values.

An additional source of heterogeneity arises from the number of individual objects (e.g., galaxies or H{\sc ii} regions) contributing to each published determination. Some studies derive $Y_\mathrm{p}$ from a single well-studied object, whereas others are based on samples comprising tens of galaxies and numerous emission-line regions. In the present analysis, each published determination is treated as a single data point, irrespective of its underlying sample size. This choice reflects our focus on the statistical behaviour of the literature as reported, rather than on reconstructing object-level measurements.

We note, however, that larger samples do not necessarily imply statistical independence or reduced systematic uncertainty, as many analyses share common modelling assumptions, calibration procedures and atomic data. Conversely, single-object studies may be more susceptible to object-specific systematics. The potential impact of these differences is therefore partially captured in the quoted uncertainties but may also contribute to the residual scatter and clustering discussed in Section~\ref{discussion.sec}. In this sense, variation in sample size represents an additional layer of methodological heterogeneity that may influence the apparent convergence of published values.

In addition, the interpretation of quoted uncertainties requires some caution. Different studies adopt heterogeneous approaches to error estimation, with some reporting statistical uncertainties only; others include partial or full treatments of systematic effects. Moreover, the definition and propagation of uncertainties have evolved over time, particularly with the increasing use of Monte Carlo and Bayesian techniques. Additionally, individual determinations are based on heterogeneous underlying samples, ranging from single objects to ensembles of tens of H{\sc ii} regions. As a result, the quoted uncertainties reflect not only measurement precision but also sample size and selection; they should not be interpreted as strictly comparable across all studies. Further, the reported error bars are not strictly homogeneous across the dataset, and inverse-variance weighting should therefore be interpreted as an approximate diagnostic rather than a formally optimal estimator.

\subsection{Tracer Classification}
\label{tracers.sec}

Each determination was assigned to a tracer category reflecting the physical system or measurement technique. Initially, these categories were: extragalactic H{\sc ii} regions, blue compact (dwarf) galaxies (BCGs), the cosmic microwave background (CMB), globular clusters, radio recombination lines, nebulae, irregular galaxies, stellar determinations and a small group classified as `other'. This scheme was used to preserve authorial intent, retaining the terminology used by the original authors wherever possible. Table~\ref{tab:overview_Yp} gives an overview of the dataset and its statistical properties by tracer. The quoted $\sigma$ values represent the unweighted standard deviation of published central values within each tracer category. Our final dataset, sorted chronologically, is available in Table \ref{tab:full_dataset} in Appendix \ref{fulldata.sec} and as an online Microsoft Excel file through \url{http://astro-expat.info/Data/pubbias.html}.\footnote{A permanent link to this page can be found at \url{http://web.archive.org/web/20200331174040/http://astro-expat.info/Data/pubbias.html}; members of the community are encouraged to send us updates or missing information.} For several tracer categories, the number of available determinations remains small. Statistical inferences based on these subsets should therefore be treated with caution.

\begin{table}
 \caption{Overview of $Y_\mathrm{p}$ determinations by tracer}
 \label{tab:overview_Yp}
 {\tablefont\begin{tabular}{@{\extracolsep{\fill}}lcccr}
  \toprule
    Tracer & $N$ & Mean $Y_\mathrm{p}$ & $\sigma$ & Skewness \\
           &     &                     &          & (post-2006)\\
  \hline
    Extragalactic H{\sc ii} regions & 56 & 0.240 & 0.011 & $-$0.733 ($-$0.726) \\
    CMB & 34 & 0.269 & 0.044 & 2.284 (2.284) \\
    BCGs & 15 & 0.233 & 0.016 & $-$1.728 \\
    Globular Clusters & 14 & 0.226 & 0.022 & $-$1.253 \\
    Radio Recombination Lines & 7 & 0.259 & 0.008 & 2.063 (1.919) \\
    Planetary Nebulae & 4 & 0.248 & 0.023 & 1.200 \\
    Irregular Galaxies & 4 & 0.231 & 0.010 & 1.706 \\
    Stars & 4 & 0.234 & 0.011 & 1.720 \\
    Others & 5 & 0.237 & 0.012 & $-$0.832 \\
  \hline
    All & 143 & 0.246 & 0.028 & 3.414
  \botrule
 \end{tabular}}
\end{table}

The categories `extragalactic H{\sc ii} regions', `irregular galaxies', and `blue compact galaxies' showed substantial overlap. Most papers in the latter two groups target low-metallicity, star-forming systems and employ the same emission-line spectroscopy, flux calibration and line-fitting pipelines as conventional H{\sc ii}-region studies. Within several tracer categories, a substantial fraction of determinations originates from a small number of author groups, a point we will revisit in Section \ref{discussion.sec} when discussing data independence and clustering. Moreover, most used identical or very similar datasets and applied the same set of systematic corrections. For this reason, these three labels were combined into a single unified tracer class, Extragalactic H{\sc ii} regions, for further analysis: see Figure \ref{fulldata.fig} (middle), where we have also included the CMB-based determinations which similarly dominate the dataset.

\subsection{From observables to primordial helium abundance}

Across all tracer classes, determinations of the primordial helium mass fraction rely on a two-step inference. First, an observable quantity---such as an emission-line flux ratio, a stellar luminosity or colour, or a feature of the CMB power spectrum---is used to estimate a present-day or early-Universe helium abundance. Second, this inferred abundance is mapped onto a primordial value, either by extrapolation to zero metallicity or by embedding the measurement within a cosmological model.

In astrophysical systems that have undergone chemical evolution, such as galaxies or stellar populations, this procedure necessarily involves assumptions about helium production beyond BBN. These assumptions are typically encoded through empirical regressions against metallicity, stellar evolution models or population synthesis calculations. In cosmological analyses, by contrast, $Y_\mathrm{p}$ enters as a parameter governing the thermal and ionisation history of the early Universe and is constrained through its effect on precision observables rather than through direct abundance measurements.

Because each step in this inference chain introduces model dependence, different tracers are subject to distinct sources of systematic uncertainty. This heterogeneity allows cross-checks between physically independent probes, but it also complicates direct comparison of published values. The following subsections summarise the principal inference mechanisms associated with the dominant tracer classes in our dataset. The two most numerous tracer groups in our dataset, extragalactic H{\sc ii} regions and the CMB, represent distinct yet complementary approaches to determining $Y_\mathrm{p}$. Although both aim to constrain the same cosmological quantity, they differ fundamentally in the physical processes and epochs they probe. Extragalactic H{\sc ii}-region measurements infer $Y_\mathrm{p}$ from the present-day Universe via spectroscopic analysis of ionised nebulae in metal-poor, star-forming galaxies, whereas CMB analyses extract it from the conditions of the early Universe, encoded in the anisotropy power spectrum. Because of their prevalence in our dataset and their contrasting methodologies, these two tracers are discussed in greater detail below.

\subsubsection{Extragalactic H{\sc ii} regions}
$Y_\mathrm{p}$ determinations from extragalactic H{\sc ii} regions are based on high-quality spectroscopic measurements of hydrogen and helium recombination lines---usually the hydrogen Balmer lines and He~{\sc i}---in metal-poor star-forming galaxies. The observed helium abundance is correlated with metallicity, typically traced by the oxygen abundance (by mass), and a linear regression to zero metallicity is performed to estimate the primordial value \citep[e.g.,][]{2007ApJ...666..636P, 2010ApJ...710L..67I, 2015JCAP...07..011A}. Analyses correct for systematic effects such as collisional excitation of helium lines, underlying stellar absorption and spatial variations in temperature and density \citep[e.g.,][]{2004ApJ...617...29O, 2012MNRAS.425L..28P}. The total uncertainty thus reflects both observational errors and modelling assumptions. Recent studies adopt homogeneous samples to minimise methodological bias, but differences in line diagnostics, temperature determinations and regression schemes continue to induce scatter (although at the sub-percent level) among published values. This approach provides a direct astrophysical estimate of the primordial helium mass fraction from the present-day Universe. 

For context, constraints on $Y_\mathrm{p}$ derived from the CMB have undergone a parallel evolution: early determinations were characterised by large uncertainties and occasionally higher central values, whereas more recent analyses based on high-precision data have converged towards $Y_\mathrm{p} \simeq 0.25$ with substantially reduced error bars. The emergence of this tight external benchmark is particularly relevant when assessing the apparent convergence of astrophysical determinations.

\subsubsection{Cosmic Microwave Background}

The CMB provides an independent and fundamentally cosmological constraint on the primordial helium abundance. The value of $Y_\mathrm{p}$ affects the recombination history of the Universe by setting the number density of free electrons prior to hydrogen recombination: a higher helium fraction reduces the electron density, thereby altering the photon diffusion length and the damping tail of the CMB temperature and polarisation power spectra. As a result, $Y_\mathrm{p}$ leaves a measurable imprint on small-scale anisotropies, partially degenerate with other cosmological parameters such as the baryon density and the effective number of relativistic species.

In modern analyses, $Y_\mathrm{p}$ is treated as either a derived quantity within the framework of standard BBN or as a free parameter in cosmological model fits to high-precision CMB data from experiments such as the Wilkinson Microwave Anisotropy Probe ({\sl WMAP}) and {\sl Planck} \citep[e.g.,][]{2016AandA...594A..13P, 2020AandA...641A...6P}. Parameter estimation is typically performed using Markov Chain Monte Carlo techniques implemented in frameworks such as \textsc{CosmoMC} \citep{2002PhRvD..66j3511L} or \textsc{MontePython} \citep{2013JCAP...02..001A}. Unlike extragalactic H{\sc ii}-region determinations, CMB-based estimates are insensitive to local astrophysical conditions and probe the early Universe directly, providing a complementary route to constraining $Y_\mathrm{p}$.

Note that a cluster of reported $Y_\mathrm{p}$ values in the early 2010s coincides with the increasing precision of cosmological predictions from {\sl WMAP} and {\sl Planck} data. During this period, the baryon density inferred from the CMB implied a predicted primordial helium fraction that observers could compare against, potentially influencing subsequently reported results. Although we do not quantify this effect here, the alignment of some published $Y_\mathrm{p}$ values with early‑2010s CMB predictions highlights the subtle interplay between observational determinations and contemporaneous theoretical benchmarks.

\subsubsection{Stellar populations and alternative astrophysical tracers}

Several determinations in our dataset rely on resolved stellar populations or integrated spectra. In globular clusters and other metal-poor stellar systems, helium abundances are inferred indirectly from stellar evolution diagnostics. These include the classical $R$ parameter \citep{1968Natur.220..143I}, as well as related indicators based on RR~Lyrae pulsation properties (the $A$ parameter), magnitude differences between evolutionary sequences \citep[e.g., the $\Delta$ parameter;][]{1983A&A...123..135C} and more recent diagnostics such as the $\xi$ parameter \citep{2011PASP..123..879T}. These methods depend sensitively on stellar evolution models and assumptions regarding age, mass loss, chemical composition and internal mixing, as well as on observational factors such as sample completeness and population selection. They typically constrain helium enrichment within stellar populations rather than providing direct measurements of primordial abundances. Conversion to $Y_\mathrm{p}$ therefore requires additional modelling to disentangle primordial helium from stellar processing and chemical evolution, and is often most robust when applied in a differential sense, for example in estimating helium-to-metal enrichment ratios \citep[e.g.,][]{2000ApJ...538..289Z}. This strong reliance on shared evolutionary models implies that different stellar-based determinations are unlikely to be fully independent, even when based on distinct observational datasets.

More recently, alternative probes such as radio recombination lines and emission from active galactic nuclei (AGN) have been explored. Radio recombination line studies infer helium abundances from centimetre-/millimetre-wavelength spectroscopy of ionised gas, offering a route that is less affected by dust extinction but subject to its own calibration challenges. AGN-based methods use the narrow-line regions of AGN as ionised gas reservoirs distinct from classical H{\sc ii} regions, providing a novel and potentially independent environment for helium abundance studies. While these approaches currently contribute relatively few determinations and often carry larger systematic uncertainties, their principal value lies in their methodological independence from the dominant extragalactic H{\sc ii}-region framework.

\subsection{Data Independence}

An additional layer of classification was applied to the extragalactic H{\sc ii} region-based determinations to assess the degree of independence among published determinations. Many studies re-analyse previously observed spectra or subsets of established samples, which can introduce correlations between nominally distinct results. To quantify this, each determination was assigned to one of three categories based on the provenance of its observational data, i.e.,
\begin{itemize}
    \item \textbf{New}: derived entirely from a newly acquired dataset not used in any earlier determination.
    \item \textbf{Partial}: based on a mixture of new and previously analysed data.
    \item \textbf{Old}: reanalysis of previously published datasets only.
\end{itemize}

This classification was carried out through a detailed examination of the data sources and observing programmes described in each source paper, cross-referenced where necessary with prior literature.
%\begin{figure}
% \includegraphics[width=\columnwidth]{Images/yp_pie.png}
% \caption{Breakdown of the compiled dataset by data independence. Each determination is labelled as new, partial or old, depending on whether it is based on an independent dataset, partially overlapping data or a re-analysis of previously published material.}
% \label{fig:yp_pie}
%\end{figure}

\begin{figure}
    \centering
    \includegraphics[width=\columnwidth]{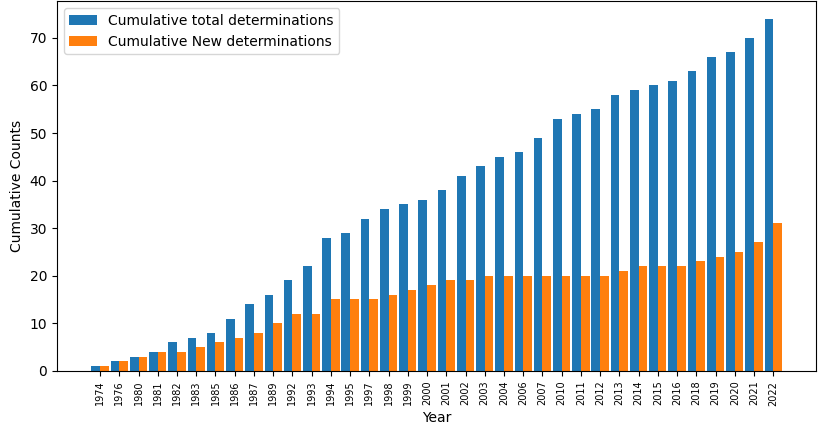}
    \vspace{0.2cm} % optional spacing between images
    \includegraphics[width=0.8\columnwidth]{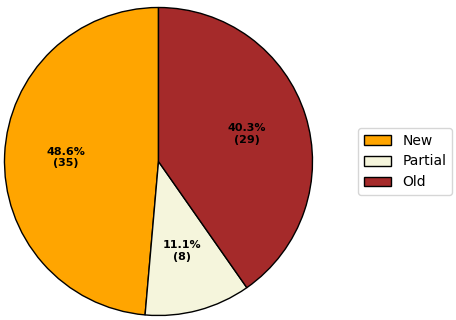}
    \caption{(Top) Cumulative number of primordial helium abundance determinations as a function of publication year. Blue bars show the total number of measurements, whereas the orange bars indicate the subset classified as `new'. The divergence between the two curves underscores the increasing contribution of re-analyses and partially overlapping data. (Bottom) Breakdown of the compilation by data independence. Each determination is labelled as `new', `partial' or `old', depending on whether it is based on an independent dataset, partially overlapping previously published data or a re-analysis of previously published material.}
    \label{fig:yp_pie}
\end{figure}

Figure~\ref{fig:yp_pie} shows the relative proportions of these categories. Approximately half of all determinations can therefore be considered fully independent, with the remainder partially or wholly dependent on earlier datasets. This breakdown provides important context for interpreting temporal and statistical patterns in our database, recognising that perfect independence is unattainable in a cumulative literature of this kind (see Section \ref{discussion.sec}).

We emphasise that independence in terms of observational data does not necessarily imply methodological independence. Many studies adopt common atomic data, emissivity calculations and analysis frameworks, which can introduce correlated systematic effects even when the underlying observations are distinct. This form of shared modelling infrastructure may contribute to the apparent convergence of results and is difficult to quantify explicitly.

\section{Methods}
\label{methods.sec}

\subsection{Author-group concentration within tracer classes}

Before turning to formal statistical tests, it is useful to characterise the structure of the published $Y_\mathrm{p}$ literature at a descriptive level. In particular, the compiled dataset reveals a pronounced concentration of published determinations within a small number of author groups, especially when considered separately by tracer.

\begin{figure*}
  \centering
 \includegraphics[width=15cm]{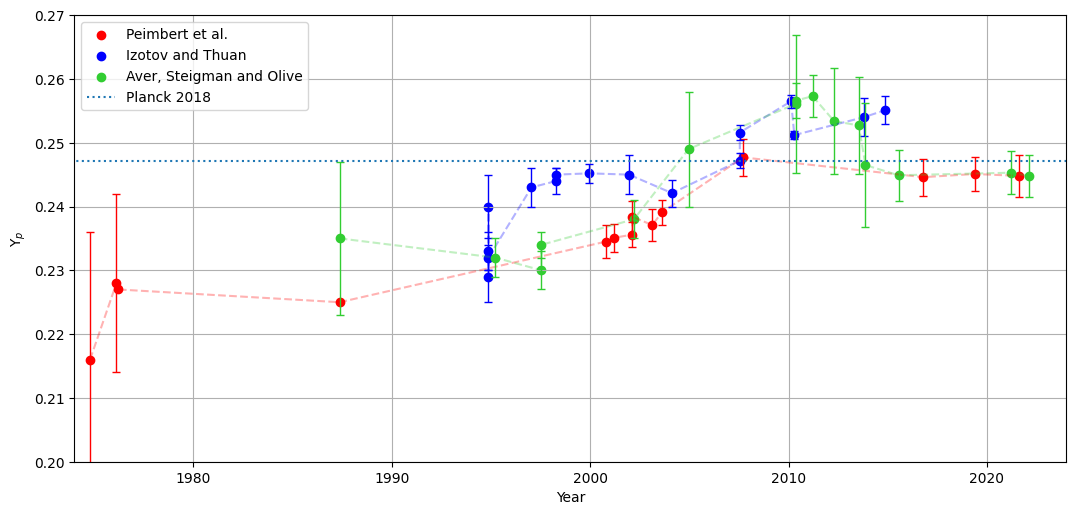}
 \caption{$Y_\mathrm{p}$ values and their uncertainties as a function of publication date published by three distinct author groups: Peimbert et al. (red), Izotov and Thuan (blue) and Aver and collaborators (green). The {\sl Planck} 2018 value, $Y_\mathrm{p} = 0.2463$ is included as a horizontal dotted line for visual benchmarking.}
 \label{3authors.fig}
\end{figure*}

Among determinations based on extragalactic H{\sc ii} regions and closely related BCG samples, more than half of all published $Y_\mathrm{p}$ values originate from three research groups: those led by Peimbert and collaborators, Olive and collaborators, and Izotov and Thuan: see Figure \ref{3authors.fig}. Collectively, these groups account for the majority of H{\sc ii}-region-based determinations over the past three decades. While these studies often differ in details of atomic data, regression schemes and systematic corrections, they frequently draw on overlapping observational samples and share a broadly similar methodological approach.

A comparable concentration is evident in other tracer classes. Of the seven determinations based on radio recombination lines, six were produced by a single group (Tsivilev and collaborators). Likewise, twelve of the fourteen globular cluster-based determinations originate from a single collaboration (Buonanno and collaborators). In contrast, CMB-based determinations exhibit a somewhat broader authorship, reflecting the collaborative nature of large survey experiments and the use of publicly released data products analysed by multiple independent teams.

These patterns are not presented here as evidence of bias or methodological deficiency. Instead, they reflect the realities of specialised observational programmes, long-term data investments and the cumulative reanalysis of high-quality datasets within relatively small expert communities. Nevertheless, such concentration has important implications for the effective independence of published determinations and for the interpretation of apparent convergence or dispersion in the literature record. In particular, a large number of nominally distinct $Y_\mathrm{p}$ values may, in practice, represent successive refinements of a limited set of underlying data and assumptions. For this reason, the author-group concentration documented here provides essential context for the statistical analyses that follow, especially those concerned with temporal convergence, variance suppression and change-point behaviour.

\subsection{Statistical Approach}
\begin{figure*}
 \includegraphics[width=8.5cm]{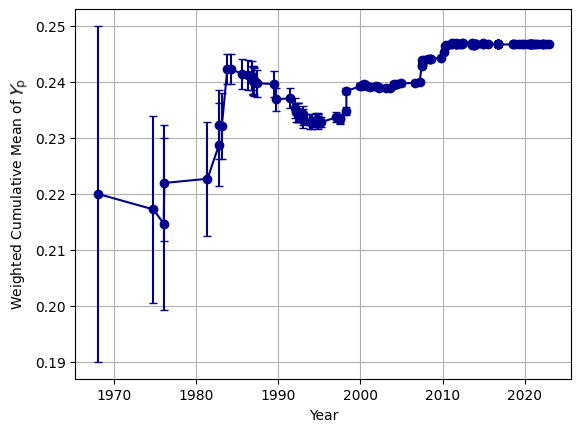}
 \includegraphics[width=8.5cm]{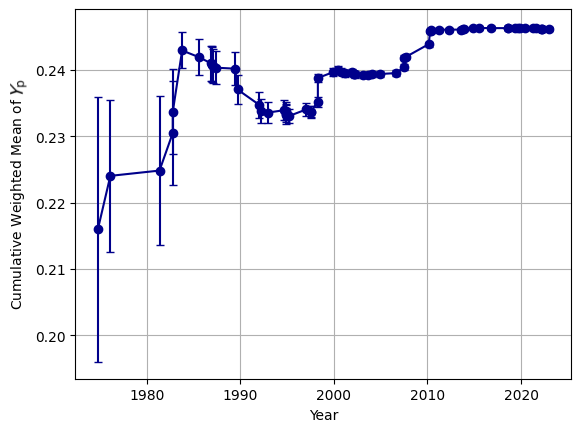}
 \caption{Weighted cumulative means as a function of publication date for (left) all published $Y_\mathrm{p}$ values and (right) $Y_\mathrm{p}$ values based on H{\sc ii} region-based analysis only.}
 \label{cumulative.fig}
\end{figure*}

Our analysis strategy was designed to probe both the distribution of reported values and their evolution over time. Three complementary strands were pursued:  
\begin{enumerate}
  \item \textbf{Descriptive statistics:} Summary statistics and distribution plots were used to characterise the overall behaviour of the dataset. Histograms of reported $Y_\mathrm{p}$ values were generated for the full sample and for selected tracer subsets. These distributions were examined for evidence of skewness, kurtosis and multi-modality, as well as for the presence of outliers that might disproportionately influence the mean. We also compared the distributions with and without CMB-based determinations to assess the impact of methodologically distinct tracers on the overall shape of the distribution. Basic summary statistics (mean, median, standard deviation) were computed for all subsets to provide a quantitative baseline for subsequent analyses.

  The inclusion of the skewness statistic in Table~\ref{tab:overview_Yp} provides an additional perspective on the distribution of published $Y_\mathrm{p}$ values within each tracer category. Whereas the standard deviation captures the overall dispersion, the skewness quantifies the degree of asymmetry in the distribution and is therefore sensitive to the presence of outliers or systematically shifted subsets of measurements. Several tracer categories exhibit non-negligible skewness, indicating that their distributions are not symmetric about the mean. In particular, the CMB-based determinations show a pronounced positive skew, driven by a subset of early high-valued estimates, whereas extragalactic H{\sc ii}-region measurements are more nearly symmetric, especially when recent determinations are considered in isolation. This behaviour is consistent with the histogram analysis presented above and reinforces the interpretation that a small number of high-valued measurements (rather than a broad underlying asymmetry) are responsible for the skewness observed in the full dataset.

  \item \textbf{Temporal trends:} To examine the historical progression of $Y_\mathrm{p}$ determinations, time-series figures were produced showing individual measurements as a function of publication date. These were constructed for the full dataset as well as for major tracer subsets (Extragalactic H{\sc ii} regions, CMB). In addition to visual inspection, we analysed the evolution of both central values and quoted uncertainties using running averages and binned statistics (see Section~\ref{clusters.sec}). Particular attention was paid to identifying periods of convergence, increased dispersion or step-like shifts in the typical inferred value. These features were then compared with known developments in observational capabilities, atomic data and cosmological constraints to assess whether changes in the literature record coincide with external scientific drivers.

  \item \textbf{Weighted means:} To assess convergence and the influence of quoted uncertainties, weighted averages of $Y_\mathrm{p}$ were computed using inverse-variance weighting. These were evaluated cumulatively (see Figure~\ref{cumulative.fig}) and in fixed and running two-year bins to trace the evolution of the inferred mean value over time. Because inverse-variance weighting can assign disproportionate influence to measurements with small quoted uncertainties, we compared these results with unweighted means and with alternative treatments that account for potential correlations between measurements (see Section~\ref{pubbias.sec}). This comparison provides a diagnostic of whether the apparent convergence is driven primarily by a subset of high-precision determinations or reflects a broader consistency across the dataset. We emphasise that these weighted means are used here as descriptive diagnostics of convergence behaviour, not as estimators of the true primordial helium abundance.
\end{enumerate}

Together, these approaches provide complementary perspectives on the distribution, temporal evolution and internal consistency of the published $Y_\mathrm{p}$ determinations.

\subsection{Change-point Analysis}
\label{cpa.sec}

To test for structural shifts in the temporal behaviour of the $Y_\mathrm{p}$ determinations, we applied change-point analysis to the extragalactic H{\sc ii}-region subset (72 data points) using the \texttt{envcpt} package in \textsf{R}. This package implements a family of likelihood-based models that allow for combinations of linear trends, discrete change points and first-order autocorrelation and is well-suited to the irregular temporal sampling characteristic of the published $Y_\mathrm{p}$ literature.

We modelled the time-series of published $Y_\mathrm{p}$ values as a function of publication year, treating the quoted measurement uncertainties as observational errors. A suite of candidate models was considered, including (i) a single linear trend with no change points; (ii) piecewise-linear models with one or more change points in the mean level and/or slope; and (iii) models incorporating a first-order autoregressive structure to account for potential temporal dependence between successive measurements. For each model, parameters were estimated via maximum-likelihood analysis.

Model selection was performed using both the Akaike and the Bayesian Information Criteria (AIC, BIC), which penalise model complexity to differing degrees. While the absolute ranking of individual models depends modestly on the chosen information criterion, both AIC and BIC consistently favour models containing multiple change points over those with a single global trend or no structural breaks. In particular, models with two change points provide a substantially better description of the data than simpler alternatives, even after accounting for the increased number of free parameters.

The preferred models identify two statistically significant change points occurring at approximately $t_1 \simeq 2006 \pm 2$ and $t_2 \simeq 2012 \pm 2$, where the quoted uncertainties reflect the width of the likelihood peaks rather than formal confidence intervals. These epochs correspond to changes in both the central tendency and the dispersion of the published $Y_\mathrm{p}$ values. Prior to the first change point, measurements exhibit a broader scatter and a lower mean value. Between the two change points, the inferred $Y_\mathrm{p}$ values increase and display reduced variance. After the second change point the distribution becomes more tightly clustered, with comparatively small quoted uncertainties.

To assess the robustness of these results, we repeated the analysis under a range of assumptions, including alternative treatments of autocorrelation, exclusion of measurements with exceptionally large or small quoted uncertainties and perturbation of individual data points. In all cases, the existence of at least two preferred change points remained stable, although their exact locations varied within the quoted uncertainties. We therefore conclude that the identification of multiple structural shifts in the published extragalactic H{\sc ii}-region $Y_\mathrm{p}$ determinations is a robust feature of the data. These epochs coincide with periods of rapid development in both observational cosmology and atomic modelling, motivating a closer examination of their broader scientific context in Section~\ref{discussion.sec}.

We emphasise that change-point analysis identifies statistically preferred structural features in the time-series but does not by itself establish causal explanations for those features. The interpretation of identified change points therefore requires external physical and historical context, which we address in the following section. Note that publication date is used here as a proxy for the temporal evolution of the field. In practice, observational data acquisition, analysis and publication can be separated by significant time intervals, and multiple studies may analyse data obtained over similar periods. This temporal smoothing may blur the precise location of change points, although it is unlikely to affect their overall identification.

\subsection{Dataset clustering and its impact on the inferred $Y_\mathrm{p}$}
\label{clusters.sec}

A further layer of structure in the extragalactic H{\sc ii}-region subset emerges when the determinations are grouped according to their underlying observational datasets. In particular, two widely used samples---hereafter denoted IT94 and IT98, following their original compilation by Izotov and Thuan---have repeatedly been re-analysed in the literature, either in isolation or combined with additional data. To assess the impact of such dataset reuse, we grouped the H{\sc ii}-region determinations into three categories: (i) those based wholly or partially on the IT94 sample (12 determinations); (ii) those based wholly or partially on the IT98 sample (20); and (iii) determinations based on datasets that appear only once in the literature (21; hereafter `singletons').

\begin{figure*}
 \includegraphics[height=6.0cm]{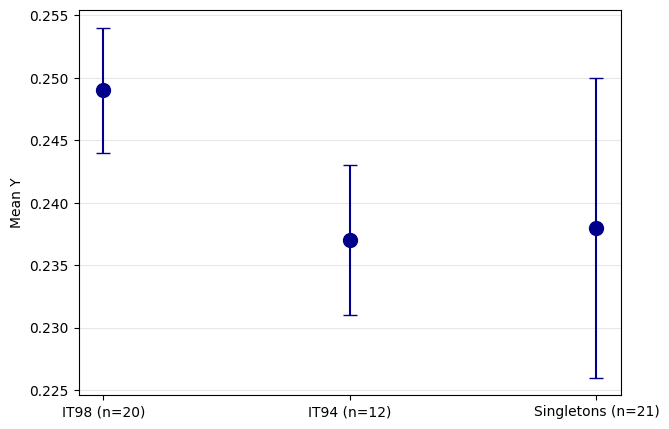}
 \includegraphics[height=6.0cm]{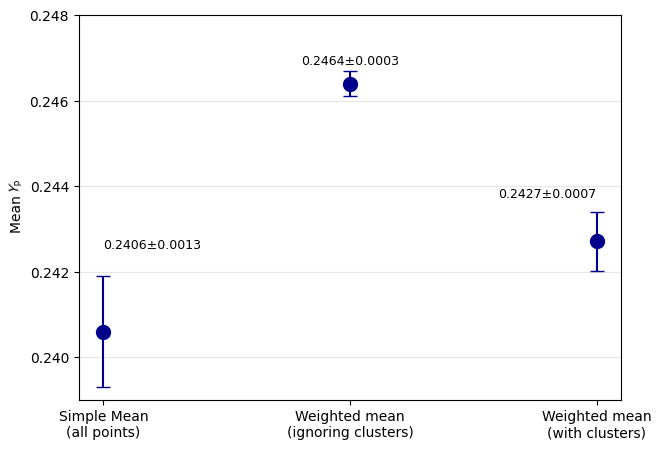}
 \caption{(left) Cluster means with their standard deviations indicated. (right) Comparison of methods.}
 \label{clusters.fig}
\end{figure*}

A comparison of the distributions of $Y_\mathrm{p}$ values across these groups reveals systematic differences: the IT98-based determinations are offset towards higher values relative to both the IT94 and singleton groups. Formal statistical tests confirm that these differences are unlikely to arise by chance: both parametric (ANOVA) and non-parametric (Kruskal--Wallis) tests reject the null hypothesis that these groups are drawn from a common parent distribution at high significance ($p < 10^{-4}$), indicating that the choice of underlying dataset has a measurable and statistically robust impact on the inferred value of $Y_\mathrm{p}$. These effects are illustrated in Figure~\ref{clusters.fig}, which compares simple, weighted and cluster-adjusted mean estimates. In the left-hand panel, we show the standard deviation within each cluster rather than the standard error on the mean, as our aim is to characterise the intrinsic spread of determinations associated with each dataset rather than the precision with which their mean value is known. Although these tests assume statistical independence and may therefore overstate formal significance in the presence of residual correlations, the magnitude of the observed differences remains striking even under conservative interpretation. 

This result demonstrates that the choice of underlying dataset can have a measurable impact on the inferred primordial helium abundance. It also highlights the extent to which repeated analysis of a small number of widely used samples can introduce structured, non-random features into the literature record.

The influence of dataset clustering is further illustrated by comparing different estimators of the mean $Y_\mathrm{p}$. A simple unweighted mean yields a value broadly representative of the overall distribution, whereas an inverse-variance weighted mean is both higher and associated with a substantially smaller formal uncertainty. This behaviour arises because a subset of determinations---primarily those associated with the IT98 sample---quote relatively small uncertainties and therefore carry disproportionate statistical weight when all measurements are treated as independent.

To account for this effect, we constructed a modified weighted mean in which the quoted uncertainties for multi-member dataset clusters are inflated by an additional term derived from the within-cluster scatter. This reduces the effective weight of highly replicated datasets while leaving singleton measurements unchanged. The resulting estimate is systematically lower than the naive weighted mean and exhibits a larger, more conservative uncertainty. Although this error-inflation scheme is necessarily approximate, it illustrates a key point: standard inverse-variance weighting can lead to biased and overconfident estimates when applied to datasets that contain significant internal correlations. A more rigorous treatment of this issue is discussed in the next section.

To provide a first consistency check of this interpretation, we compared the cluster-adjusted mean described above with the expectation from a simple hierarchical model in which a non-zero inter-dataset dispersion is allowed. Even without performing a full parameter inference, the observed shift between the naive inverse-variance weighted mean and the cluster-adjusted estimate is consistent with the presence of a finite between-dataset variance, $\tau$. This supports the view that the dispersion between datasets is non-negligible and should be accounted for when combining measurements. This behaviour illustrates that even when the underlying observational material is held fixed, methodological choices alone can produce a dispersion comparable to that seen across nominally independent studies.

The apparent level of agreement between clusters depends sensitively on how uncertainties are represented. When standard errors on the mean are used, clusters with many determinations appear artificially precise, exaggerating the significance of inter-cluster differences. By contrast, representations based on within-cluster dispersion reveal that a substantial component of the scatter arises from dataset-level systematics, consistent with the hierarchical structure of the compiled sample.

\section{Discussion}
\label{discussion.sec}

\subsection{Long-term trends and convergence in published $Y_\mathrm{p}$ values}
\label{trends.sec}

The most striking feature of the $Y_\mathrm{p}$ dataset is the pronounced evolution in both the central values and the quoted uncertainties over time. Early determinations exhibit substantial scatter, with inferred primordial helium abundances spanning a wide range and often accompanied by large statistical and systematic uncertainties. In contrast, measurements published since the mid-2000s display a clear narrowing of the distribution, with values increasingly concentrated within a comparatively small interval. We stress that the observed convergence of $Y_\mathrm{p}$ determinations does not, by itself, imply any pathology in the literature. Improved data quality, better atomic physics and more consistent analysis pipelines naturally lead to reduced scatter over time. The question addressed here is not whether convergence should occur but whether its timing, rapidity and cross-method coherence are fully accounted for by these factors alone.

Convergence is evident across multiple statistical representations of the data, including time-series plots, running means and histograms. Whereas the full dataset exhibits mild asymmetry driven by a small number of high-valued outliers---primarily associated with early CMB-based determinations---the distribution of non-CMB measurements alone is approximately symmetric about its mean. This suggests that the apparent skewness of the full sample is not an intrinsic feature of observational helium abundance studies but instead reflects the inclusion of methodologically distinct tracers with different systematic characteristics.

The temporal evolution of the quoted uncertainties is similarly revealing. Prior to approximately 2005, error bars are typically large and heterogeneous, reflecting both observational limitations and an incomplete understanding of key systematic effects. In subsequent years, uncertainties decrease markedly, and by the past decade many determinations quote errors at the few-percent level or better. This trend is consistent with improvements in spectroscopic data quality, atomic physics inputs and analysis techniques, but it also raises the question of whether residual systematics are uniformly assessed across studies.

When considered separately by tracer, additional structure emerges. Extragalactic H{\sc ii}-region determinations dominate the dataset numerically and show the clearest convergence behaviour; measurements based on globular clusters, variable stars and other stellar tracers remain comparatively sparse and more widely dispersed. CMB-based determinations, although fewer in number, occupy a distinct region of parameter space. They evolve in parallel with improvements in cosmological data and modelling.

The skewness statistics reported in Table~\ref{tab:overview_Yp} further support this interpretation. The reduction in skewness over time for the dominant tracer categories suggests that the apparent convergence is not only a narrowing of the distribution but also a suppression of asymmetry, consistent with the decreasing influence of early outliers and the increasing homogenisation of analysis methods. Although for tracer categories with small sample sizes, skewness values should be interpreted with caution, the presence of skewness may also reflect the influence of clustered measurements derived from common underlying datasets, as discussed in Section~\ref{pubbias.sec}.

Importantly, the observed convergence is not monotonic. Periods of relative stability are punctuated by shifts in the typical inferred value of $Y_\mathrm{p}$, as quantified by our change-point analysis in Section~\ref{cpa.sec}. These shifts suggest that convergence has occurred through a series of distinct adjustments rather than through smooth, incremental refinement. Understanding the drivers of these adjustments is essential for interpreting the present consensus and is explored in greater detail in the following subsections. We emphasise that convergence in itself is not evidence of (publication) bias. In a mature field, improved data quality, refined methodologies and better understanding of systematic effects are expected to produce a narrowing of the inferred parameter range. The key question is therefore whether the rate and structure of the expected convergence are consistent with methodological progress alone.

\subsection{Change points and external cosmological context}

Although the number of data points used for our change-point analysis in Section~\ref{cpa.sec} (72) is modest, our application of two independent statistical approaches (AIC and BIC) consistently identifies two statistically preferred epochs, centred approximately on the years 2006 and 2012, when the temporal behaviour of published extragalactic H{\sc ii}-region determinations of $Y_\mathrm{p}$ exhibits significant shifts. These change points are evident both in the evolution of individual measurements and in the behaviour of cumulative and binned means, and they are robust across a range of model assumptions and information criteria.

The earlier change point occurs shortly after the publication of \citet{2004ApJ...617...29O}, which introduced a revised treatment of systematic effects in H{\sc ii}-region analyses and yielded a primordial helium abundance notably higher than most earlier determinations. In the years following this analysis, published $Y_\mathrm{p}$ values show an overall upward trend and increased clustering around values near $Y_\mathrm{p} \simeq 0.25$. This period also coincides with the release of the first high-precision CMB constraints on the baryon density from \textit{WMAP}, which for the first time enabled a quantitatively precise comparison between BBN-based abundance determinations and cosmological measurements derived from the early Universe.

A second prominent change point occurs during the period 2012--2015, coinciding with a sequence of methodological revisions rather than a single defining publication. In particular, the introduction of updated He~{\sc i} emissivities by \citet{2012MNRAS.425L..28P, 2013MNRAS.433L..89P}, together with their subsequent adoption in extragalactic H{\sc ii}-region analyses \citep[e.g.,][]{2013AandA...558A..57I, 2014MNRAS.445..778I, 2015JCAP...07..011A}, led to systematic downward revisions of inferred helium abundances and a reduction in internal scatter. This phase marks the emergence of a comparatively narrow range of published $Y_\mathrm{p}$ values, typically clustering around $Y_\mathrm{p} \simeq 0.245$--0.248, with smaller quoted uncertainties than in earlier work. At approximately the same time, the increasing use of the near-infrared He~{\sc i} $\lambda$10,830{\AA} transition provided a powerful new diagnostic of H{\sc ii}-region physical conditions, particularly electron density, thereby helping to reduce degeneracies in nebular modelling and improve constraints on systematic uncertainties. This line rapidly became a standard component of modern helium abundance analyses and likely contributed significantly to the tightening of the post-2012 distribution of published $Y_\mathrm{p}$ values.

This consolidation coincides temporally with the availability of high-precision CMB-based baryon density measurements from {\sl Planck}, which yield an inferred primordial helium abundance in close agreement with the revised extragalactic determinations. This marks the beginning of a period extending for nearly a decade during which published $Y_\mathrm{p}$ values cluster tightly within a narrow interval, with comparatively small quoted uncertainties. Notably, this behaviour persists across multiple studies employing different datasets and observational facilities.

The most recent change in the time series is associated with the EMPRESS\footnote{Extremely Metal-Poor Representatives Explored by the Subaru Survey} determination of \citet{2022ApJ...941..167M}, which yields a value of $Y_\mathrm{p}$ substantially lower than the prevailing cluster of earlier measurements. Unlike the previous two change points, this result represents a departure from the CMB-inferred value, not a convergence towards it. This highlights that deviations from the apparent consensus can and do still occur within the published literature. However, more recent analyses by \citet{2026arXiv260122238A} recover values consistent with the primordial helium abundance expected from the combination of {\sl Planck} cosmology, the Standard Model of particle physics and standard BBN calculations, thereby reducing the apparent tension introduced by the EMPRESS result. This illustrates that the field continues to dynamically evolve and that apparent departures from consensus values continue to be tested and reassessed through new observations and revised analyses.

These change points underscore the importance of situating $Y_\mathrm{p}$ determinations within their broader context. The temporal alignment between shifts in extragalactic H{\sc ii}-region measurements and major advances in CMB cosmology does not, by itself, imply causal influence. Nevertheless, it suggests that the interpretation and presentation of observational results may be shaped, at least in part, by evolving external benchmarks and expectations regarding concordance among independent cosmological probes. Distinguishing between genuine methodological progress and more subtle forms of anchoring or convergence therefore requires careful consideration of both statistical patterns and the historical context in which measurements are produced.

\subsection{Independence, convergence and the question of publication bias}
\label{pubbias.sec}

The identification of statistically significant change points and the marked convergence of published $Y_\mathrm{p}$ values in the past two decades naturally raises the question of whether this behaviour reflects genuine methodological progress alone, or whether sociological effects---including various forms of publication bias---may also play a role. By `publication bias' we do not imply deliberate data manipulation or selective reporting but rather the more subtle tendency for subsequent analyses to converge on widely cited benchmark values, either through shared assumptions, common calibration choices or the implicit weighting of earlier high-profile results.

The dataset-level analysis presented in Section~\ref{clusters.sec} provides further insight into the origin of this convergence. The existence of statistically significant offsets between determinations based on different underlying samples demonstrates that the literature is not sampling a single homogeneous distribution of $Y_\mathrm{p}$ values. Instead, it reflects a superposition of partially correlated measurement clusters, each tied to a specific observational dataset and analysis tradition.

In this context, the apparent precision of inverse-variance weighted averages must be interpreted with caution. Because certain datasets---notably the IT98 sample---have been re-analysed multiple times with comparatively small quoted uncertainties, they exert disproportionate influence on weighted estimates. As a result, naive combinations of the literature can yield values that are both shifted relative to the broader distribution and associated with artificially small uncertainties.

A key difficulty in assessing such effects lies in the independence of the published determinations. As discussed in Section~\ref{methods.sec}, a substantial fraction of extragalactic H{\sc ii}-region measurements are not fully independent, in the sense that their authors re-analyse overlapping datasets with modified atomic data, revised emissivities or alternative treatments of systematic uncertainties. However, even when attention is restricted to the subset of determinations classified as fully independent---those based on entirely new observational datasets---a striking degree of convergence remains evident, particularly in the period after $\sim$2012.

A closer examination of independent extragalactic H{\sc ii}-region studies provides further insight into the evolution and apparent convergence of published $Y_\mathrm{p}$ values. In particular, six independent determinations published between 2018 and 2022, based on wholly distinct datasets and observational facilities, yield values of $Y_\mathrm{p}$ clustered tightly around 0.245--0.247. These studies encompass a variety of observational material, including Sloan Digital Sky Survey \citep[SDSS;][]{2018MNRAS.478.5301F} samples, Very Large Telescope spectroscopy \citep[e.g.,][]{2019ApJ...876...98V, 2021MNRAS.505.3624V} and Keck/Large Binocular Telescope observations \citep[e.g.,][]{2020ApJ...896...77H, 2021JCAP...03..027A, 2022MNRAS.510..373A}. They employ independent analysis pipelines and atomic emissivity corrections. Despite their methodological diversity, the inferred helium abundances exhibit remarkable agreement, suggesting that modern determinations have reached a level of consistency---including with that inferred from CMB analyses---that may reflect both improved data quality and convergent treatment of systematic uncertainties. 

Moreover, \citet{2022MNRAS.514.5506D} used SDSS spectra of AGN to determine their $Y_\mathrm{p}$ value, a novel method appearing in literature for the first time. This is also a highly independent determination, since it uses an entirely new class of objects. The fact that this clustering persists even among formally independent datasets suggests that convergence is not driven solely by data reuse, but it may also reflect shared modelling assumptions, common calibration choices or implicit anchoring to externally favoured values. The probability that six genuinely independent analyses, based on distinct datasets and methodologies, would converge to this degree purely by chance is therefore likely small, reinforcing the need to consider shared modelling assumptions or external anchoring as contributing factors.

This convergence is, on the one hand, encouraging and may reasonably be interpreted as evidence that the field has matured, with improved atomic data, better control of systematic effects and higher-quality observations leading to a stable and reliable estimate of $Y_\mathrm{p}$. On the other hand, the tight clustering of results from ostensibly independent analyses also invites closer scrutiny. In particular, it raises the question whether shared modelling choices, prior expectations informed by CMB results or the widespread adoption of similar atomic emissivities and correction schemes may effectively reduce the true diversity of the determinations, even when the underlying observational data are distinct.

By contrast, earlier determinations, particularly the influential study of \citet{1998ApJ...500..188I}, demonstrated the challenges of the field in the late 1990s. These authors employed high signal-to-noise spectroscopy of a modest sample of low-metallicity H{\sc ii} regions, applying detailed corrections for underlying stellar absorption, collisional excitation and temperature structure. Whereas their reported uncertainty was small for the period, subsequent reanalyses and methodological updates have revised the inferred $Y_\mathrm{p}$ upwards by a few percent, highlighting both the sensitivity of helium determinations to systematic assumptions and the evolving treatment of atomic data.

The temporal context is important. The second change point identified in Section~\ref{cpa.sec} occurs shortly after the release of the first {\sl Planck} cosmological results, which provided a high-precision, model-dependent prediction for the primordial helium abundance. Whereas extragalactic H{\sc ii}-region analyses and CMB measurements are based on fundamentally different physical probes, they are not entirely insulated from one another. CMB-based values can influence the choice of priors, the perceived plausibility of particular systematic corrections or the interpretation of outlying results in later observational studies. The visual similarity between the temporal evolution of CMB- and H{\sc ii}-region-based $Y_\mathrm{p}$ determinations is therefore striking, even if it does not, by itself, constitute evidence of bias.

Placing these two epochs---\citet{1998ApJ...500..188I} and the recent independent studies---side-by-side illustrates the dual nature of convergence in the literature. On the one hand, modern determinations reflect methodological improvements, homogeneous adoption of updated emissivities and higher-quality observational datasets. On the other hand, the clustering of recent independent results around a narrow interval, in close agreement with CMB-inferred predictions, also underscores the potential influence of widely cited benchmarks and shared methodological frameworks. Recognising this interplay between genuine methodological refinement and sociological effects is crucial for interpreting the present consensus and for assessing the effective independence of published $Y_\mathrm{p}$ values.

Further insight is provided by the behaviour of determinations based on alternative tracers. Methods relying on Galactic globular clusters, RR~Lyrae stars or, more recently, AGN spectra exhibit a smaller number of published measurements and are often dominated by a single research group. Although this limits their statistical weight, it also highlights the extent to which certain tracers have not been subjected to the same level of independent replication as extragalactic H{\sc ii}-region studies. The emergence of new, genuinely distinct probes of $Y_\mathrm{p}$ therefore remains an important avenue for testing the robustness of the apparent convergence.

Finally, the most recent determination in our compilation departs noticeably from the clustered values of the preceding decade, suggesting that convergence is neither inevitable nor irreversible. Whether such deviations reflect improved treatment of previously underestimated systematics, genuinely new physical insights or simply statistical fluctuation cannot yet be determined. However, their presence underscores the importance of maintaining methodological diversity and of critically examining consensus values, particularly in a field where theoretical expectations are both precise and influential.

In summary, these considerations suggest that the observed convergence in published $Y_\mathrm{p}$ values is likely driven by a combination of genuine methodological advances and more subtle sociological effects. Distinguishing between these contributions is inherently challenging, but the patterns identified here closely parallel those found in other areas of astrophysics where publication bias has been investigated. In the next section, we discuss possible strategies for mitigating such effects and for strengthening future empirical constraints on the primordial helium abundance.

\subsection{Implications and paths forward}
\label{implications.sec}

The results presented in this paper have implications both for the interpretation of existing primordial helium abundance determinations and for the design of future studies. Whereas the overall convergence of published $Y_\mathrm{p}$ values over the past two decades may reflect genuine progress in observational quality and atomic modelling, the statistical patterns identified here suggest that caution is warranted when interpreting the apparent precision of the current consensus.

A central implication of our analysis is that the effective information content of the published literature may be lower than the raw number of quoted measurements would suggest. The widespread reuse of a small number of observational datasets, particularly those pertaining to extragalactic H{\sc ii}-region studies, and the adoption of common modelling assumptions and atomic data, imply that many published determinations are not statistically independent. As a result, naive averaging or meta-analysis that treats all measurements on an equal footing risks underestimating systematic uncertainties and overestimating the significance of apparent agreement.

Future compilations of primordial helium abundance measurements would benefit from a more explicit treatment of dependence and correlation. One practical step would be the routine classification of determinations according to their degree of data reuse, distinguishing between genuinely new observations, partial re-analyses and full re-analyses of existing datasets. Such classification would enable weighted analyses that more accurately reflect the true diversity of the underlying information and would help to prevent overrepresentation of heavily reused datasets.

More broadly, methodological diversity should be actively encouraged. Independent observational probes that rely on fundamentally different physical environments or diagnostics---such as stellar populations, variable stars or AGN---play a crucial role in testing the robustness of extragalactic H{\sc ii}-region results. Although such methods currently contribute fewer measurements and often larger uncertainties, their value lies precisely in their independence from the dominant techniques and assumptions. Continued development of these alternative tracers is therefore not just useful but essential.

The interaction between observational determinations of $Y_\mathrm{p}$ and high-precision cosmological predictions derived from the CMB also deserves careful consideration. Although consistency between these approaches is a powerful validation of the standard cosmological model, it also creates the potential for subtle feedback between theory and observation. In recognition of this, recent cosmological analyses have increasingly embraced formal blinding methodologies or pre-defined analysis procedures to mitigate the influence of prior expectations on parameter inference. For example, blinding transformations applied at the summary‑statistic level have been developed for multiprobe cosmological experiments \citep{2020MNRAS.494.4454M}, and practical blinding strategies have been implemented in major survey analyses such as the Dark Energy Survey Year-3 weak lensing and clustering programme \citep{2025MNRAS.536.1303J}. Examples of modern cosmological analyses that integrate high‑precision CMB data with independent large‑scale structure surveys illustrate the level of statistical precision now attainable, while also highlighting the importance of rigorous analysis protocols including blinding \citep{2025JCAP...10..077D}. Such strategies, in which key cosmological benchmarks are withheld until observational pipelines are finalised, may therefore help to reduce the influence of prior expectations on published results and ensure that independent datasets contribute genuinely independent information. This temporal coincidence is particularly striking given that the two approaches rely on fundamentally different data and physical modelling, suggesting that convergence may arise not only from improved measurements but also from shared expectations of concordance.

Finally, our results underscore the importance of reporting and propagating systematic uncertainties in a transparent and conservative manner. The historical evolution of quoted error bars suggests that improvements in precision have not always been accompanied by a corresponding increase in acknowledged sources of uncertainty. Explicit exploration of model dependence, alternative atomic datasets and analysis choices should therefore be viewed not as a weakness but as an essential component of robust analyses.

In summary, the determination of the primordial helium abundance remains a cornerstone of observational cosmology. Ensuring that future measurements are both precise and demonstrably independent will be crucial for fully exploiting its diagnostic power and for avoiding the pitfalls of premature consensus. Taken together, the results of the change-point analysis (Section~\ref{cpa.sec}) and the dataset-level clustering analysis point to a consistent picture. The temporal shifts identified in the literature coincide not only with methodological developments but also with periods during which particular datasets and analysis frameworks gained prominence. The hierarchical modelling presented below and in Appendix \ref{appendix:hierarchical} further demonstrates that treating such clusters as independent measurements can bias both the inferred central value and its uncertainty. These findings therefore suggest that the observed convergence in $Y_\mathrm{p}$ is shaped by a combination of genuine physical insight, shared data resources and methodological standardisation, rather than by independent replication alone.

\subsection*{Hierarchical treatment of dataset-level correlations}

The results presented in Section~\ref{clusters.sec} highlight a fundamental limitation of standard approaches to combining published $Y_\mathrm{p}$ determinations: measurements are often treated as statistically independent even when they are derived from the same underlying observational datasets. A more principled treatment can be achieved using a hierarchical (multi-level) model, in which correlations between measurements are explicitly represented.

Although we provide a more in-depth discussion in Appendix \ref{appendix:hierarchical}, in its simplest form we may write the $i$-th determination of $Y_\mathrm{p}$ as
\begin{equation}
    Y_i \sim \mathcal{N}(\theta_{c[i]}, \sigma_i^2),
\end{equation}
where $\sigma_i$ is the quoted measurement uncertainty and $\theta_{c[i]}$ is the latent mean associated with the dataset (or cluster) $c$ to which measurement $i$ belongs. The dataset-level parameters are themselves drawn from a global distribution,
\begin{equation}
    \theta_c \sim \mathcal{N}(\mu, \tau^2),
\end{equation}
where $\mu$ represents the overall mean primordial helium abundance and $\tau$ quantifies the intrinsic dispersion between independent datasets.

In this framework, repeated analyses of the same dataset contribute information about $\theta_c$ but do not artificially increase the effective weight of that dataset in constraining $\mu$. The parameter $\tau$ naturally captures inter-dataset variability arising from systematic differences in data quality, analysis choices or physical environments. This formulation provides a direct statistical interpretation of the behaviour observed in Section~\ref{clusters.sec}: the naive inverse-variance weighted mean corresponds to the limiting case $\tau \rightarrow 0$, in which all datasets are assumed to be perfectly consistent, whereas the cluster-adjusted estimates implicitly allow for a non-zero between-dataset variance ($\tau > 0$; see Appendix~\ref{appendix:hierarchical}). As a result, hierarchical models yield more conservative and, arguably, more realistic uncertainties on $\mu$ when dataset-level correlations are present.

A full implementation of this approach, including parameter inference and model comparison, is left to future work. However, the simple formulation outlined here provides a natural framework for interpreting the structured variability observed in the $Y_\mathrm{p}$ literature and for constructing statistically robust combined estimates. This formulation is equivalent to a random-effects meta-analysis in which between-cluster variance is explicitly modelled, providing a principled alternative to ad hoc error inflation schemes.

\section{Conclusions}
\label{concl.sec}

We have presented a comprehensive literature-based analysis of published determinations of the primordial helium mass fraction, $Y_\mathrm{p}$, spanning more than five decades of observational and cosmological research. By compiling and classifying essentially all observationally derived $Y_\mathrm{p}$ values reported between the late-1960s and 2022, we have constructed a dataset that enables both an historical perspective on the field and a quantitative assessment of its statistical properties.

Our analysis confirms a pronounced convergence of published $Y_\mathrm{p}$ values in recent decades, accompanied by a substantial reduction in quoted uncertainties. While this trend is broadly consistent with genuine advances in observational capabilities, atomic physics inputs and analysis techniques, we find that the evolution of the literature is structured rather than smooth. Change-point analysis identifies statistically significant epochs when the typical inferred value of $Y_\mathrm{p}$ shifts, suggesting discrete adjustments in methodology or interpretation rather than continuous refinement.

A key finding of this study is the limited degree of independence among published determinations, particularly for extragalactic H{\sc ii}-region measurements. A substantial fraction of the literature is based on re-analyses or partial reuse of a small number of observational datasets, often coupled with closely related modelling assumptions. This interdependence reduces the effective number of independent constraints on $Y_\mathrm{p}$ and provides important context for interpreting the apparent agreement among recent measurements. We emphasise that our results do not call into question the internal consistency of the standard cosmological model or the broad agreement between astrophysical and CMB-based determinations of the primordial helium abundance. Instead, they highlight the need for careful interpretation of precision claims and for explicit consideration of correlation, methodological homogeneity and historical context when synthesising results across the literature.

Future progress will depend not only on improved data quality but also on methodological diversity, transparent treatment of systematics and the continued development of genuinely independent observational probes. In this sense, the case of $Y_\mathrm{p}$ illustrates how apparent precision in a mature field can emerge not only from improved data and methods but also from the structured evolution of the literature itself. In combination, these steps will ensure that constraints on $Y_\mathrm{p}$ remain both precise and robust, preserving its central role as a probe of early-Universe physics. More broadly, the patterns identified here highlight the importance of treating cumulative measurement literatures as evolving systems, in which statistical, methodological and sociological factors may all contribute to the emergence of consensus values.

\section*{Acknowledgements}

We are grateful to Prof. Georgy Sofronov (Macquarie University, Sydney, Australia) for his advice on our statistical analysis using the \texttt{envcpt} package in \textsf{R}. We thank the referee for their constructive report and particularly for highlighting the important role of the near-infrared He~{\sc i} $\lambda$10,830{\AA} transition in the evolution of modern H{\sc ii}-region analyses. This research has made extensive use of the Astrophysics Data System, funded by NASA under Cooperative Agreement 80NSSC21M00561.

\section*{Data Availability}

This paper is a meta-analysis of published values for the primordial helium abundance. A reference to our full compilation is provided in Section \ref{tracers.sec}.

\bibliographystyle{mnras}
\bibliography{pubbias} 

%%%%%%%%%%%%%%%%% APPENDICES %%%%%%%%%%%%%%%%%%%%%

\appendix
\renewcommand{\thetable}{A\arabic{table}}
\setcounter{table}{0}

\onecolumn
\section{Full dataset}
\label{fulldata.sec}

Table \ref{tab:full_dataset} contains details of our full dataset. CMB-based determinations are flagged according to the methodology used to obtain $Y_\mathrm{p}$:

\noindent
$^1$ $Y_\mathrm{p}$ is treated as a free parameter and constrained directly from a fit to the CMB anisotropy power spectrum;

\noindent
$^2$ The baryon density, $\Omega_{\rm b} h^2$, is constrained from the CMB fit and subsequently used to infer $Y_\mathrm{p}$ under the assumptions of standard $\Lambda$CDM cosmology and BBN.

\setlength{\tabcolsep}{20pt}
\renewcommand{\arraystretch}{1.5}

\begin{longtable}{l c l c l}
\caption{Compilation of primordial helium abundance determinations. For a description of the CMB flags, see the text.}
\label{tab:full_dataset} \\

\hline
Month/Year & $Y_{\mathrm{p}} \pm \sigma$ & Tracer & Indep.\ ID & Reference \\
\hline
\endfirsthead

\multicolumn{5}{c}{\tablename\ \thetable\ -- continued from previous page} \\
\hline
Year & $Y_{\mathrm{p}} \pm \sigma$ & Tracer & Data Type & Reference \\
\hline
\endhead

\hline
\multicolumn{5}{r}{Continued on next page} \\
\endfoot

\hline
\endlastfoot

01/1968 & $0.22 \pm 0.03$ & Miscellaneous & New & \citet{1968ApL.....2...91B} \\
06/1969 & $0.26$ & RRL & New & \citet{1969ApJ...156..887P}  \\
10/1974 & $0.216 \pm 0.02$ & H\,\textsc{ii} regions & New & \citet{1974ApJ...193..327P} \\
01/1976 & $0.20 \pm 0.04$ & GC & New & \citet{1976PhDT.........3T} \\
02/1976 & $0.228 \pm 0.014$ & H\,\textsc{ii} regions & New & \citet{1976ApJ...203..581P} \\
03/1976 & $0.227$ & Nebulae & New & \citet{1976AandA....47..341D} \\
04/1977 & $0.28$ & Nebulae & New & \citet{1977ApJ...213..421B} \\
08/1980 & $0.216$ & H\,\textsc{ii} regions & New & \citet{1980ApJ...240...41F} \\
05/1981 & $0.24 \pm 0.05$  & H\,\textsc{ii} regions & New & \citet{1981ApJ...246...38T} \\
08/1982 & $0.23$ & Star & New & \citet{1982PASP...94..634D} \\
10/1982 & $0.236 \pm 0.011$ & H\,\textsc{ii} regions & Old & \citet{1982RSPTA.307...37K} \\
10/1982 & $0.24 \pm 0.011$ & H\,\textsc{ii} regions & Old & \citet{1982RSPTA.307...37K} \\
02/1983 & $0.23 \pm 0.02$ & GC & New & \citet{1983prhe.work..231B} \\
10/1983 & $0.245 \pm 0.003$ & H\,\textsc{ii} regions & New & \citet{1983ApJ...273...81K} \\
03/1984 & $0.25 \pm 0.25$ & Star & New & \citet{1984AJ.....89..433M} \\
07/1985 & $0.212 \pm 0.015$ & BCG & New & \citet{1985ApJS...58..321D} \\
04/1986 & $0.23 \pm 0.02$ & GC & New & \citet{1986AandA...159..189B} \\
10/1986 & $0.24 \pm 0.01$ & Miscellaneous & Old & \citet{1986PASP...98..984K}  \\
11/1986 & $0.207 \pm 0.016$ & H\,\textsc{ii} regions & Old & \citet{1986ApJ...310L..67F} \\
12/1986 & $0.23 \pm 0.02$ & H\,\textsc{ii} regions & New & \citet{1986ApJ...311...45D} \\
12/1986 & $0.25 \pm 0.02$ & H\,\textsc{ii} regions & New & \citet{1986ApJ...311...45D} \\
01/1987 & $0.23 \pm 0.07$ & GC & New & \citet{1987AandAS...67..327B} \\
01/1987 & $0.19 \pm 0.03$ & BCG & Old & \citet{1987AandA...172...15V} \\
05/1987 & $0.235 \pm 0.012$ & H\,\textsc{ii} regions & Old & \citet{1987RMxAA..14...71S} \\
05/1987 & $0.225$ & H\,\textsc{ii} regions & New & \citet{1987RMxAA..14..540P} \\
06/1989 & $0.234 \pm 0.016$ & BCG & New & \citet{1989AJ.....97.1591D} \\
09/1989 & $0.229 \pm 0.004$ & H\,\textsc{ii} regions & New & \citet{1989RMxAA..18..153P} \\
06/1991 & $0.238 \pm 0.004$ & Nebulae & New & \citet{1991ApJ...374..580B} \\
10/1991 & $0.23$ & GC & New & \citet{1991MNRAS.252..357F} \\
01/1992 & $0.216 \pm 0.006$ & BCG & New & \citet{1992AandA...253...16M} \\
03/1992 & $0.228 \pm 0.005$ & H\,\textsc{ii} regions & New & \citet{1992MNRAS.255..325P} \\
06/1992 & $0.23 \pm 0.05$ & Star & New & \citet{1992ApJ...392..172H} \\
06/1992 & $0.24 \pm 0.05$ & GC & New & \citet{1992MNRAS.256..376F} \\
06/1992 & $0.25 \pm 0.05$ & GC & New & \citet{1992MNRAS.256..376F} \\
12/1992 & $0.24$ & H\,\textsc{ii} regions & Old & \citet{1992ApJ...401..157C} \\
01/1993 & $0.228 \pm 0.005$ & Irregular galaxy & Old & \citet{1993ApJ...403...65M} \\
01/1993 & $0.223 \pm 0.006$ & Irregular galaxy & Old & \citet{1993ApJ...403...65M} \\
11/1993 & $0.227 \pm 0.006$ & Irregular galaxy & Old & \citet{1993ApJ...418..229B} \\
01/1994 & $0.22$ & H\,\textsc{ii} regions & New & \citet{1994ApJ...420..576M} \\
02/1994 & $0.23 \pm 0.04$ & GC & New & \citet{1994MNRAS.266..829F} \\
08/1994 & $0.239 \pm 0.006$ & BCG & New & \citet{1994ApJ...431..172S} \\
10/1994 & $0.23 \pm 0.02$ & GC & New & \citet{1994AandA...290...69B} \\
11/1994 & $0.229 \pm 0.004$ & BCG & New & \citet{1994ApJ...435..647I} \\
11/1994 & $0.240 \pm 0.005$ & BCG & New & \citet{1994ApJ...435..647I} \\
11/1994 & $0.233 \pm 0.003$ & BCG & New & \citet{1994ApJ...435..647I} \\
11/1994 & $0.232 \pm 0.002$ & BCG & New & \citet{1994ApJ...435..647I} \\
03/1995 & $0.232 \pm 0.003$ & H\,\textsc{ii} regions & Old & \citet{1995ApJS...97...49O} \\
06/1996 & $0.233$ & Miscellaneous & New & \citet{1996AJ....111.2135R} \\
01/1997 & $0.243 \pm 0.003$ & H\,\textsc{ii} regions & Partial & \citet{1997ApJS..108....1I} \\
07/1997 & $0.230 \pm 0.003$ & H\,\textsc{ii} regions & Partial & \citet{1997ApJ...483..788O} \\
07/1997 & $0.234 \pm 0.002$ & H\,\textsc{ii} regions & Partial & \citet{1997ApJ...483..788O} \\
04/1998 & $0.244 \pm 0.002$ & BCG & New & \citet{1998SSRv...84...83T} \\
04/1998 & $0.245 \pm 0.001$ & BCG & New & \citet{1998SSRv...84...83T} \\
12/1999 & $0.2452 \pm 0.0015$ & BCG & New & \citet{1999ApJ...527..757I} \\
03/2000 & $0.225 \pm 0.013$ & Star & New & \citet{2000MNRAS.313...99R} \\
06/2000 & $0.2489 \pm 0.0030$ & Miscellaneous & Old & \citet{2000ApJ...536..773B} \\
10/2000 & $0.2345 \pm 0.0026$ & H\,\textsc{ii} regions & New & \citet{2000ApJ...541..688P} \\
03/2001 & $0.2351 \pm 0.0022$ & H\,\textsc{ii} regions & Old & \citet{2001RMxAC..10..148P} \\
12/2001 & $0.245 \pm 0.003$ & BCG & New & \citet{2001ApJ...562..727I} \\
02/2002 & $0.2356 \pm 0.0020$ & H\,\textsc{ii} regions & Old & \citet{2002ApJ...565..668P} \\
02/2002 & $0.2384 \pm 0.0025$ & H\,\textsc{ii} regions & Old & \citet{2002ApJ...565..668P} \\
03/2002 & $0.238 \pm 0.003$ & H\,\textsc{ii} regions & Old & \citet{2002ApJ...567..931G} \\
02/2003 & $0.2371 \pm 0.0025$ & H\,\textsc{ii} regions & New & \citet{2003ApJ...584..735P} \\
08/2003 & $0.2391 \pm 0.0020$ & H\,\textsc{ii} regions & Old & \citet{2003ApJ...592..846L} \\
01/2004 & $0.2574 \pm 0.0025$ & RRL & New & \citet{2004OAP....17..103T} \\
02/2004 & $0.2421 \pm 0.0021$ & BCG & Partial & \citet{2004ApJ...602..200I} \\
06/2004 & $0.250 \pm 0.006$ & GC & New & \citet{2004AandA...420..911S} \\
12/2004 & $0.249 \pm 0.009$ & H\,\textsc{ii} regions & Old & \citet{2004ApJ...617...29O} \\
08/2006 & $0.250 \pm 0.004$ & H\,\textsc{ii} regions & Old & \citet{2006ApJ...646..691F} \\
03/2007 & $0.252 \pm 0.004$ & RRL & Partial & \citet{2007AandA...464..995P} \\
06/2007 & $0.24819^{+0.00029}_{-0.00060}$ & CMB$^2$ & New & \citet{2007ApJS..170..377S} \\
07/2007 & $0.2472 \pm 0.0012$ & H\,\textsc{ii} regions & Old & \citet{2007ApJ...662...15I} \\
07/2007 & $0.2516 \pm 0.0011$ & H\,\textsc{ii} regions & Old & \citet{2007ApJ...662...15I} \\
09/2007 & $0.2477 \pm 0.0029$ & H\,\textsc{ii} regions & Old & \citet{2007ApJ...666..636P} \\
04/2008 & $0.2479 \pm 0.0014$ & Nebulae & Old & \citet{2008ARep...52..327H} \\
06/2008 & $0.2486 \pm 0.0085$ & CMB$^1$ & Partial & \citet{2008JCAP...06..028P} \\
06/2008 & $0.2487^{+0.0451}_{-0.0484}$ & CMB$^1$ & Partial & \citet{2008JCAP...06..028P} \\
08/2008 & $0.25^{+0.10}_{-0.07}$ & CMB$^1$ & Old & \citet{2008PhRvD..78d3509I} \\
02/2009 & $0.45$ & CMB$^1$ & Old & \citet{2009ApJS..180..306D} \\
10/2009 & $0.254 \pm 0.002$ & RRL & Old & \citet{2009AstL...35..670T} \\
02/2010 & $0.2565 \pm 0.0010$ & H\,\textsc{ii} regions & Old & \citet{2010ApJ...710L..67I} \\
04/2010 & $0.2512 \pm 0.0006$ & H\,\textsc{ii} regions & Partial & \citet{2010IAUS..268..107I} \\
05/2010 & $0.2561 \pm 0.0108$ & H\,\textsc{ii} regions & Old & \citet{2010JCAP...05..003A} \\
05/2010 & $0.2566 \pm 0.0028$ & H\,\textsc{ii} regions & Old & \citet{2010JCAP...05..003A} \\
01/2011 & $0.2584 \pm 0.0018$ & RRL & New & \citet{2011OAP....24...43M} \\
02/2011 & $0.326 \pm 0.075$ & CMB$^1$ & Partial & \citet{2011ApJS..192...14J} \\
03/2011 & $0.2573 \pm 0.0033$ & H\,\textsc{ii} regions & Old & \citet{2011JCAP...03..043A} \\
08/2011 & $0.181 \pm 0.014$ & GCs & Partial & \citet{2011PASP..123..879T} \\
08/2011 & $0.183 \pm 0.014$ & GCs & Partial & \citet{2011PASP..123..879T} \\
08/2011 & $0.238 \pm 0.014$ & GCs & Partial & \citet{2011PASP..123..879T} \\
08/2011 & $0.236 \pm 0.014$ & GCs & Partial & \citet{2011PASP..123..879T} \\
09/2011 & $0.313 \pm 0.044$ & CMB$^1$ & Partial & \citet{2011ApJ...739...52D} \\
12/2011 & $0.296 \pm 0.030$ & CMB$^1$ & Partial & \citet{2011ApJ...743...28K} \\
04/2012 & $0.2534 \pm 0.0083$ & H\,\textsc{ii} regions & Old & \citet{2012JCAP...04..004A} \\
05/2012 & $0.303 \pm 0.075$ & CMB$^1$ & Old & \citet{2012PhRvD..85j3519G} \\
05/2012 & $0.277 \pm 0.050$ & CMB$^1$ & Old & \citet{2012PhRvD..85j3519G} \\
06/2012 & $0.279 \pm 0.028$ & CMB$^1$ & Partial & \citet{2012AstL...38..347B} \\
07/2013 & $0.2527 \pm 0.0076$ & H\,\textsc{ii} regions & New & \citet{2013AJ....146....3S} \\
07/2013 & $0.2463 \pm 0.0003$ & CMB$^2$ & New & \citet{2013arXiv1307.6955C} \\
10/2013 & $0.299 \pm 0.027$ & CMB$^1$ & Old & \citet{2013ApJS..208...19H} \\
10/2013 & $0.254 \pm 0.003$ & H\,\textsc{ii} regions & Partial & \citet{2013AandA...558A..57I} \\
10/2013 & $0.225 \pm 0.034$ & CMB$^2$ & Partial & \citet{2013JCAP...10..060S} \\
11/2013 & $0.2564 \pm 0.0018$ & RRL & New & \citet{2013AstL...39..737T} \\
11/2013 & $0.2465 \pm 0.0097$ & H\,\textsc{ii} regions & Old & \citet{2013JCAP...11..017A} \\
12/2013 & $0.298 \pm 0.031$ & CMB$^1$ & Partial & \citet{2013PDU.....2..188L} \\
02/2014 & $0.305 \pm 0.024$ & CMB$^1$ & Partial & \citet{2014ApJ...782...74H} \\
02/2014 & $0.300 \pm 0.025$ & CMB$^1$ & Partial & \citet{2014ApJ...782...74H} \\
11/2014 & $0.266 \pm 0.021$ & CMB$^1$ & New & \citet{2014AandA...571A..16P} \\
11/2014 & $0.2551 \pm 0.0022$ & H\,\textsc{ii} regions & New & \citet{2014MNRAS.445..778I} \\
11/2014 & $0.3194 \pm 0.0130$ & CMB$^1$ & Old & \citet{2014AandA...571A..22P} \\
07/2015 & $0.2449 \pm 0.0040$ & H\,\textsc{ii} regions & Old & \citet{2015JCAP...07..011A} \\
09/2016 & $0.24665^{+0.00020}_{-0.00019}$ & CMB$^2$ & Partial & \citet{2016AandA...594A..13P} \\
09/2016 & $0.24667 \pm 0.00018$ & CMB$^2$ & Partial & \citet{2016AandA...594A..13P} \\
09/2016 & $0.24667 \pm 0.00014$ & CMB$^2$ & Partial & \citet{2016AandA...594A..13P} \\
09/2016 & $0.24668 \pm 0.00013$ & CMB$^2$ & Partial & \citet{2016AandA...594A..13P} \\
09/2016 & $0.253^{+0.041}_{-0.042}$ & CMB$^1$ & Partial & \citet{2016AandA...594A..13P} \\
09/2016 & $0.255^{+0.036}_{-0.038}$ & CMB$^1$ & Partial & \citet{2016AandA...594A..13P} \\
09/2016 & $0.251^{+0.026}_{-0.027}$ & CMB$^1$ & Partial & \citet{2016AandA...594A..13P} \\
09/2016 & $0.253^{+0.025}_{-0.026}$ & CMB$^1$ & Partial & \citet{2016AandA...594A..13P} \\
10/2016 & $0.2446 \pm 0.0029$ & H\,\textsc{ii} regions & Old & \citet{2016RMxAA..52..419P} \\
08/2018 & $0.244 \pm 0.006$ & H\,\textsc{ii} regions & New & \citet{2018MNRAS.478.5301F} \\
08/2018 & $0.245 \pm 0.007$ & H\,\textsc{ii} regions & New & \citet{2018MNRAS.478.5301F} \\
10/2018 & $0.250^{+0.033}_{-0.025}$ & Miscellaneous & New & \citet{2018NatAs...2..957C} \\
05/2019 & $0.2451 \pm 0.0026$ & H\,\textsc{ii} regions & New & \citet{2019ApJ...876...98V} \\
08/2019 & $0.243 \pm 0.005$ & H\,\textsc{ii} regions & Old & \citet{2019MNRAS.487.3221F} \\
11/2019 & $0.2468 \pm 0.0032$ & H\,\textsc{ii} regions & Partial & \citet{2019JPhCS1400b2051K} \\
01/2020 & $0.277$ & RRL & New & \citet{2020OAP....33...18T} \\
06/2020 & $0.2436 \pm 0.0040$ & H\,\textsc{ii} regions & New & \citet{2020ApJ...896...77H} \\
09/2020 & $0.24672^{+0.00011}_{-0.00012}$ & CMB$^2$ & Partial & \citet{2020AandA...641A...6P} \\
09/2020 & $0.24714^{+0.00012}_{-0.00013}$ & CMB$^2$ & Partial & \citet{2020AandA...641A...6P} \\
09/2020 & $0.241 \pm 0.025$ & CMB$^1$ & Partial & \citet{2020AandA...641A...6P} \\
09/2020 & $0.243^{+0.023}_{-0.024}$ & CMB$^1$ & Partial & \citet{2020AandA...641A...6P} \\
12/2020 & $0.211 \pm 0.031$ & CMB$^1$ & Partial & \citet{2020JCAP...12..047A} \\
12/2020 & $0.220 \pm 0.018$ & CMB$^1$ & Partial & \citet{2020JCAP...12..047A} \\
12/2020 & $0.232 \pm 0.011$ & CMB$^1$ & Partial & \citet{2020JCAP...12..047A} \\
03/2021 & $0.2453 \pm 0.0034$ & Irregular galaxy & New & \citet{2021JCAP...03..027A} \\
04/2021 & $0.2462 \pm 0.0022$ & H\,\textsc{ii} regions & Partial & \citet{2021MNRAS.502.3045K} \\
08/2021 & $0.2448 \pm 0.0033$ & H\,\textsc{ii} regions & New & \citet{2021MNRAS.505.3624V} \\
02/2022 & $0.2448 \pm 0.0033$ & BCG & New & \citet{2022MNRAS.510..373A} \\
03/2022 & $0.229 \pm 0.014$ & H\,\textsc{ii} regions & New & \citet{2022MNRAS.510.4436M} \\
03/2022 & $0.236 \pm 0.012$ & H\,\textsc{ii} regions & New & \citet{2022MNRAS.510.4436M} \\
08/2022 & $0.2441 \pm 0.0037$ & Miscellaneous & New & \citet{2022MNRAS.514.5506D} \\
12/2022 & $0.2370^{+0.0033}_{-0.0034}$ & H\,\textsc{ii} regions & New & \citet{2022ApJ...941..167M} \\

\end{longtable}
\twocolumn

\section{Hierarchical modelling of $Y_\mathrm{p}$ determinations}
\label{appendix:hierarchical}

Here we outline a simple hierarchical framework for combining published determinations of the primordial helium abundance in the presence of dataset-level correlations.

\subsection{Model specification}

Let $Y_i$ denote the $i$-th published estimate of $Y_\mathrm{p}$ with quoted uncertainty $\sigma_i$. Each measurement is associated with a dataset (or cluster) index $c[i] \in \{1, \dots, C\}$, where $C$ is the total number of distinct underlying datasets.

We model the measurements as
\begin{equation}
    Y_i \mid \theta_{c[i]} \sim \mathcal{N}(\theta_{c[i]}, \sigma_i^2),
\end{equation}
where $\theta_c$ is the latent true value associated with dataset $c$. This level captures the fact that multiple measurements based on the same dataset are not independent but instead scatter around a common dataset-specific value.

At the next level, the dataset parameters are assumed to be drawn from a global distribution,
\begin{equation}
    \theta_c \mid \mu, \tau \sim \mathcal{N}(\mu, \tau^2),
\end{equation}
where $\mu$ represents the overall primordial helium abundance and $\tau$ quantifies the dispersion between datasets. The parameter $\tau$ may be interpreted as an effective measure of inter-dataset systematic variability.

\subsection{Interpretation}

This model separates two distinct sources of variability:
\begin{itemize}
    \item \textit{Within-dataset scatter}, governed by the quoted measurement uncertainties $\sigma_i$;
    \item \textit{Between-dataset scatter}, governed by $\tau$.
\end{itemize}

In the limit $\tau \rightarrow 0$, all dataset means collapse to a single value, $\theta_c = \mu$, and the model reduces to a standard inverse-variance weighted mean. Conversely, for $\tau > 0$, measurements from the same dataset contribute less independent information about $\mu$, preventing heavily re-analysed datasets from dominating the combined estimate.

\subsection{Likelihood and inference}

Marginalising over the latent dataset parameters $\{\theta_c\}$ yields a likelihood for the global parameters $(\mu, \tau)$,
\begin{equation}
    p(\{Y_i\} \mid \mu, \tau) = \prod_{c=1}^{C} 
    \int \left[ \prod_{i \in c} \mathcal{N}(Y_i \mid \theta_c, \sigma_i^2) \right]
    \mathcal{N}(\theta_c \mid \mu, \tau^2) \, {\rm d}\theta_c.
\end{equation}

This integral has an analytic solution, yielding an effective likelihood in which each dataset contributes according to both its internal scatter and the between-dataset variance, $\tau^2$.

Parameter estimation may be performed using maximum likelihood or Bayesian methods. In a Bayesian framework, weakly informative priors may be adopted, for example,
\begin{equation}
    \mu \sim \mathcal{N}(0.25, 0.05^2), \quad \tau \sim \mathrm{Half\text{-}Cauchy}(0, 0.02),
\end{equation}
although the precise choice of priors has little impact, provided they are broad relative to the scale of the data.

\subsection{Analytic marginal likelihood}

The integral over the latent dataset parameters $\theta_c$ can be evaluated in closed form. For a given dataset (cluster) $c$ containing measurements $\{Y_i\}$ with uncertainties $\{\sigma_i\}$, define the inverse-variance weights
\begin{equation}
    w_i = \sigma_i^{-2}, \qquad W_c = \sum_{i \in c} w_i,
\end{equation}
and the weighted mean
\begin{equation}
    \bar{Y}_c = \frac{1}{W_c} \sum_{i \in c} w_i Y_i.
\end{equation}

Marginalising over $\theta_c$ yields an effective likelihood for the cluster,
\begin{equation}
    p(\{Y_i\}_{i \in c} \mid \mu, \tau) 
    = \mathcal{N}\!\left(\bar{Y}_c \mid \mu, \, \tau^2 + W_c^{-1}\right)
    \times \prod_{i \in c} \mathcal{N}\!\left(Y_i \mid \bar{Y}_c, \sigma_i^2\right).
\end{equation}

The second term depends only on the internal scatter within the cluster and is independent of $\mu$ and $\tau$. For inference on the global parameters, the relevant contribution is therefore
\begin{equation}
    p(\bar{Y}_c \mid \mu, \tau) = \mathcal{N}\!\left(\bar{Y}_c \mid \mu, \, \tau^2 + W_c^{-1}\right).
\end{equation}

The full likelihood becomes a product over clusters,
\begin{equation}
    p(\{Y_i\} \mid \mu, \tau) \propto \prod_{c=1}^{C} 
    \mathcal{N}\!\left(\bar{Y}_c \mid \mu, \, \tau^2 + W_c^{-1}\right).
\end{equation}

\subsection{Connection to cluster-adjusted estimators}

The structure of the marginal likelihood provides a direct interpretation of the empirical cluster-based weighting scheme explored in Section~\ref{clusters.sec}. In that approach, the uncertainties of measurements belonging to the same dataset are inflated by a term derived from the within-cluster scatter. In the hierarchical framework, this role is played by the parameter $\tau$, which represents the intrinsic dispersion between datasets.

The effective variance associated with each dataset,
\begin{equation}
    \sigma_{c,\mathrm{eff}}^2 = \tau^2 + W_c^{-1},
\end{equation}
naturally combines measurement uncertainty (through $W_c^{-1}$) with inter-dataset variability (through $\tau^2$). In this sense, the hierarchical model can be viewed as a principled generalisation of the ad hoc error inflation procedure, with $\tau$ inferred directly from the data rather than imposed externally. This formalism therefore provides a principled statistical underpinning for the empirical cluster-based corrections explored in the main text and offers a natural path towards more robust meta-analyses of $Y_\mathrm{p}$ determinations.

%%%%%%%%%%%%%%%%%%%%%%%%%%%%%%%%%%%%%%%%%%%%%%%%%%
\end{document}